\documentclass[11pt,a4paper]{article}
\usepackage{jheppub}
\usepackage[utf8]{inputenc}
\usepackage[english]{babel}

\usepackage{amsfonts}
\usepackage{diagbox}

\usepackage{color, colortbl}

\newcommand\RR[1]{\multicolumn{1}{c|}{#1}}
\newcommand\LR[1]{\multicolumn{1}{|c|}{#1}}
\newcommand{\eps}{\varepsilon}
\newcommand{\dx}{\mathrm{d}}
\newcommand{\iu}{\mathrm{i}}

\newcommand{\iunderbrace}[1]{\underbrace{\scriptstyle#1}}
\newcommand{\clausen}{\mathrm{Cl}_2}
\newcommand{\pochhammer}[2]{\left(#1\right)_{#2}}
\newcommand{\polylog}{\mathrm{Li}}
\newcommand{\goncharov}{\mathrm{G}}

\title{On multi-propagator angular integrals}

\author[a]{Juliane Haug,}
\author[b,c]{Vladimir A. Smirnov,}
\author[a]{and Fabian Wunder}

\affiliation[a]{Institute for Theoretical Physics, University of T\"ubingen, \\ 
	Auf der Morgenstelle 14, 72076 T\"ubingen, Germany}
\affiliation[b]{Skobeltsyn Institute of Nuclear Physics, Moscow State University, \\ Leninskie gory, 119992 Moscow, Russia}
\affiliation[c]{Moscow Center for Fundamental and Applied Mathematics 119992 Moscow, Russia}

\emailAdd{juliane-clara-celine.haug@uni-tuebingen.de}
\emailAdd{smirnov@theory.sinp.msu.ru}
\emailAdd{fabian.wunder@uni-tuebingen.de}

\abstract{}
\arxivnumber{}

\abstract{
We study multi-propagator angular integrals, a class of phase-space integrals relevant to processes with multiple observed final states and a test-bed for transferring loop-integral technology to phase-space integrals without reversed unitarity.
We present an Euler integral representation similar to the Lee-Pomeransky representation and explicitly 
describe a recursive IBP reduction and dimensional shift relations for the general case of $n$ denominators.
On the level of master integrals, applying a differential equation approach, we explicitly calculate the previously unknown angular integrals with four denominators for any number of masses to finite order in $\varepsilon$.
Extending the idea of dimensional recurrence, we explore the decomposition of angular integrals into branch integrals reducing the number of scales in the master integrals from $(n+1)n/2$ to $n+1$.
To showcase the potential of this method, we calculate the massless three denominator integral and establish all-order results in $\varepsilon$, including a resummation of soft logarithms.}

\begin{document}
\maketitle

\flushbottom

\section{Introduction}
\label{sec: Introduction}
Angular integrals \cite{Ellis:1980,Schellekens:1981, vanNeerven:1985, Beenakker:1988, Somogyi:2011, Lyubovitskij:2021, Wunder:2024,Smirnov:2024pbj} are an integral part of phase-space integrals in perturbative quantum field theory calculations.\footnote{After publication of the preprint, the authors were made aware that a class of (hyperspherical) angular integrals has also been used for loop integration in \citep{Laporta:1994mb}.
By Wick rotating Minkowski space, these angular integrals are formulated in even spatial dimensions $d$, while their phase-space counterparts studied in this work are in odd (spatial) dimensions $d-1$. The dimensional shift relations of sec.\,\ref{sec: Dimensional shift relation} only relate $d\rightarrow d\pm 2$, so these two classes of integrals are not connected by them.
Nevertheless, there are striking similarities, e.g. between eq.\,(15) of \citep{Laporta:1994mb} and eq.\,(8.19) of \citep{Haug:2024yfi} hinting at a potential (yet unknown) relation $d\rightarrow d\pm1$.
}
In the quantum chromodynamics (QCD) literature, they were, for example, used in theoretical predictions for deep-inelastic scattering (DIS) \cite{Duke:1982,Hekhorn:2019}, semi-inclusive deep-inelastic scattering (SIDIS) \cite{Anderle:2016,Wang:2019,Ahmed:2024owh}, the Drell-Yan process (DY) \cite{Matsuura:1989, Matsuura:1990,Hamberg:1991,Mirkes:1992,Bahjat-Abbas:2018}, hadron-hadron scattering \cite{Ellis:1980}, heavy quark production \cite{Beenakker:1988}, prompt-photon production \cite{Gordon:1993,Rein:2024} and single-spin asymmetries \cite{Schlegel:2012,Ringer:2015,Rein:2025qhe,Rein:2025pwu}.
Further recent applications of angular integrals can be found in \cite{Devoto:2024,Rowe:2024jml,Agarwal:2024gws,Baranowski:2024ysi,Baranowski:2025}; similar angular integrals in the context of cosmology arose in \cite{Lee:2024nqu}.

Generically, calculations with massless particles lead to collinear divergences, which are regularized by performing the calculation in $d= 4-2\eps$ dimensions \cite{tHooft:1972, Bollini:1972}.
On the one hand, non-integer dimensionality causes these phase-space integrals to be non-trivial and requires careful analytic treatment. 
On the other hand, we will see that introducing a dependence on the space time dimension as a new variable opens up a rich structure within this class of integrals to be discovered.

Angular integrals are classified by the number of denominators (i.e. propagators) and masses (i.e. non-light-like vectors).
The more denominators and masses, the more complex the integral.
The literature mentioned above only required integrals with up to two denominators because partial fraction decomposition could be used to reduce higher-denominator integrals.
This was possible due to the restricted kinematics with a total of only three observed particles in initial and final states.
Going to more exclusive processes with a larger number of observed particles will require genuine multi-propagator integrals that cannot be reduced by partial fraction decomposition, because of less restricted kinematics with a larger number of linearly independent momenta.

Recently, there has been a resurgence of interest in angular integrals.
For the two-denominator case, where analytic results in $d$ dimensions have been known for some time \cite{vanNeerven:1985,Beenakker:1988,Somogyi:2011,Lyubovitskij:2021}, $\eps$-expansions to all orders, for an arbitrary number of masses, were given in \cite{Lyubovitskij:2021}.
The small-mass asymptotic behavior of two-denominator integrals was examined in \cite{Wunder:2024}.
Expanding on  that work, multi-denominator integrals in the limit of small masses were studied 
in \cite{Smirnov:2024pbj} using Euler representations \cite{Salvatori:2024nva} 
and expansion by regions \cite{Beneke:1997zp,Smirnov:2002pj,Smirnov:2012gma}, where $\eps$-expansions up to finite order and to leading power in the masses were given for the three-denominator integral with an arbitrary number of masses.
In addition, the collinear pole structure for the general case of $n$ denominators has been conjectured in this study.
In \cite{Haug:2024yfi}, three-denominator angular integrals with integer denominator powers and an arbitrary number of masses were expanded up to order $\eps$.
Using integration-by-parts identities, mass reduction, a dimensional shift identity and differential equations, compact results in terms of Clausen functions as well as a handful of logarithms and dilogarithms were established.
A more cumbersome version of the same results, including a lengthy expression for the order $\eps^2$ of the massless three-denominator integral, was simultaneously given in \cite{Ahmed:2024pxr} with a calculation based on a Mellin-Barnes representation.
Furthermore, the single-massive four-denominator angular integral has been calculated to order $\eps^0$ as an example of a novel approach based on the tropical geometry in \cite{Salvatori:2024nva}, additionally showcasing an interest in angular integrals. Within this approach, a given Euler integral can be represented as a linear
combination of integrals (called {\em locally finite}), where an expansion in $\eps$ under integral sign is possible.  

In this paper we build on this knowledge in two ways.
First, we can present several findings regarding the general structure for an arbitrary number of denominators.
This includes an Euler integral representation similar to Lee-Pomeransky representation, explicitly recursive integration-by-parts (IBP) reduction, dimensional shift relations, and a scale reduction in terms of \textit{branch integrals} to reduce the scale of master integrals.
Second, we provide explicit new results for three- and four-denominator angular integrals by calculating the three-denominator massless integral to all orders and the four-denominator integral to finite order for any number of masses.
Table \ref{tab: Overview} gives an overview about the $\eps$-expansions known in the literature and the new additions.

\begin{table}
\centering
\scalebox{0.9}{
\begin{tabular}{c|ccccc}
\backslashbox{\# denom.}{\# masses} & 0 & 1 & 2 & 3 & 4 \\
\hline
0 &  $\infty^+$\,\cite{Ellis:1980} & -- & -- & -- & -- \\
1 &  $\infty^+$\,\cite{Ellis:1980} & $\infty^+$\,\cite{Beenakker:1988,Somogyi:2011,Lyubovitskij:2021,Wunder:2024} & -- & -- & -- \\
2 &  $\infty^+$\,\cite{Ellis:1980,vanNeerven:1985,Lyubovitskij:2021} & $\infty^+$\,\cite{Beenakker:1988,Somogyi:2011,Lyubovitskij:2021,Wunder:2024} & $\infty^+$\,\cite{Schellekens:1981,Lyubovitskij:2021,Wunder:2024} & -- & -- \\
\cline{1-6}
\LR{3} & 2\,\cite{Haug:2024yfi,Ahmed:2024pxr}\,$\rightarrow$\,\textcolor{red}{$\infty^+$} & $1$\,\cite{Haug:2024yfi,Ahmed:2024pxr} & $1$\,\cite{Haug:2024yfi,Ahmed:2024pxr} & $1$\,\cite{Haug:2024yfi,Ahmed:2024pxr}& \RR{--}\\
\LR{4} &  $-1$\,\cite{Smirnov:2024pbj}\,$\rightarrow$\,\textcolor{red}{$0$} & $0$\,\cite{Smirnov:2024pbj,Salvatori:2024nva}\,$\rightarrow$\,\textcolor{red}{$0$} & $-1$\,\cite{Smirnov:2024pbj}\,$\rightarrow$\,\textcolor{red}{$0$} &  $-1$\,\cite{Smirnov:2024pbj}\,$\rightarrow$\,\textcolor{red}{$0$} & \RR{$-1$\,\cite{Smirnov:2024pbj}\,$\rightarrow$\,\textcolor{red}{$0$}} \\
\cline{1-6}
\dots & \dots & \dots & \dots & \dots & \dots \\
$n$ &  $-1^*$\,\cite{Smirnov:2024pbj} & $-1^*$\,\cite{Smirnov:2024pbj} & $-1^*$\,\cite{Smirnov:2024pbj} & $-1^*$\,\cite{Smirnov:2024pbj} & $-1^*$\,\cite{Smirnov:2024pbj}\\
\end{tabular}
}
\label{tab: Overview}
\caption{Overview of known orders in the $\eps$-expansion of angular integrals and what is added in this paper (highlighted in red).
A superscript ``$+$" indicates an expansion including resummation of soft logarithms; superscript ``$*$'' marks a conjectured result.
The listed references are the most significant contributions towards the state-of-the-art of the respective integrals. 
For the massive three-denominator integrals higher orders in $\eps$ can be constructed from the branch integral $\mathcal{B}_3^{(1)}$ given in Appendix \ref{app: Discussion of specific branch integrals}.
For the single-massive four-denominator integral we give a new, more compact form compared to the original calculation from \cite{Salvatori:2024nva}.}
\end{table}

To calculate the $\eps$-expansion of the four-denominator angular integral, 
our first approach was based on the so-called 
two-point splitting lemma \cite{Wunder:2024} which enables us to write down a product of two
massive factors in angular integrals as a linear combination of two terms with one massless and one massive factor each.
Then a given four-denominator angular integral with four non-zero masses can be presented as a linear
combination of four-denominator integrals with only one non-zero mass. For the various resulting
integrals with one non-zero mass, we then applied the analytic result by Salvatori
\cite{Salvatori:2024nva} and obtained an analytic result for the general four-denominator angular integral
with four non-zero masses. Proceeding in a similar way, we obtained analytic results for integrals with 
three and two non-zero masses. 
In fact, depending on the order of application of the two-point splitting lemma,
we obtained different versions of analytic results to order $\eps^0$. However, they all turned out to be rather
cumbersome so that we switched to our second approach which provided an explicit expansion of the four-denominator angular integral up to finite order $\eps^0$, with compact results in terms of Clausen functions.
This latter method we used is described in detail in ref.\,\cite{Haug:2024yfi}, a graphical overview is given in figure 1 of this reference. Here, the main tool are IBP relations and differential equations aided by a dimensional shift identity.

The all-order $\eps$-expansion of the massless three-denominator integral is obtained by applying dimensional recurrence.
This method was originally proposed in the context of loop integrals \cite{Tarasov:1996br}, where it was found that hypergeometric functions appearing in solutions of recurrence relations with respect to $d$ have fewer arguments than the results found using other methods \cite{Tarasov:2000sf, Fleischer:2003rm, Davydychev:1990jt, Davydychev:1990cq, Anastasiou:1999ui, Tarasov:2019mqy}.
From the all-order expansion we finally recover the soft singularity structure and resum the large logarithms to all orders in $\eps$ resulting in an all-order expansion that is well behaved in the soft limit.

Following the notation of references \cite{Somogyi:2011,Lyubovitskij:2021,Smirnov:2024pbj}, the general angular integral is given by
	\begin{align}
		I_{j_1,j_2, \ldots, j_n}^{(m)} \left(v_1,\ldots v_n, \eps \right) \,=\, \int \frac{\dx \Omega_{d-1}(k)}{\Omega_{d-3}}  \prod_{i=1}^n \frac{1}{(v_i\cdot k)^{j_i}} \,,
	\end{align}
with spherically symmetric integration measure
\begin{align}
\dx\Omega_{d-1}(k) = \prod_{i=1}^{n}\dx\theta_i\,\sin^{d-2-i}\theta_i\,\dx\Omega_{d-1-n}(k)\,,
\end{align}
and normalized $d$-vectors 
	\begin{align}
		v_i \,=\, (1,\boldsymbol{v_i}) \,, \quad k\,=\, (1,\boldsymbol{k}) \,=\,(1,\ldots, \cos\theta_n \prod_{i=1}^{n-1} \sin\theta_i, \ldots, \cos\theta_2\sin\theta_1,\cos\theta_1) \,.
	\end{align}
The integral depends on the invariants $v_i\cdot v_j\equiv v_{ij}$.
The superscript $m$ characterizes the number of non-zero masses $v_{ii} = v_i \cdot v_i$, while the denominator powers $j_i$ are assumed to be integers in the following.
The normalization factor $ 1/\Omega_{d-3}$ simplifies the $\eps$-expansion by removing factors of the Euler-Mascheroni constant $\gamma_E$.
For the unnormalized angular integrals there is the notation (see \cite{Somogyi:2011})
\begin{align}
\Omega_{j_1,\dots,j_n}(\lbrace v_{kl}\rbrace;\eps)=\Omega_{d-3}\,I_{j_1,j_2, \ldots, j_n}^{(m)} \left(v_1,\ldots v_n, \eps \right)
\end{align}
we will use in sec.\,\ref{sec: Euler representation for angular integral}.

At this point we want to briefly mention that these angular integrals are close analogues of one-loop Feynman integrals. 
In a pioneering paper \cite{vanNeerven:1985} the two-denominator angular integral has been calculated from the absorptive part of a box integral.
In \cite{Somogyi:2011}, \cite{Lyubovitskij:2021}, and \cite{Haug:2024yfi} direct methods proved effective to extend these results. Also in this work we will directly transfer loop-integral technology to angular integrals without using reversed unitarity \cite{Anastasiou:2002,Anastasiou:2003,Anastasiou:2004}.
Akin to the box to two-denominator angular integral correspondence, a curious reader may compare the three- and four-denominator angular integrals discussed in this work to the $\eps$-expansion of the pentagon \cite{Kozlov:2016} and hexagon \cite{Henn:2022ydo} Feynman integrals.

The remainder of this paper is organized as follows.
In Section \ref{sec: Euler representation for angular integral}
we present an Euler integral representation similar to Lee-Pomeransky representation, 
improving a representation given in \cite{Smirnov:2024pbj}.
Section \ref{sec: IBP relations} deals with IBP relations in an explicit recursive form that allow for a reduction of angular integrals to master integrals without the need for Laporta's algorithm \cite{Laporta:2000}.
In Section \ref{sec: Dimensional shift relation}, we state the dimensional shift identity for four propagators, and conjecture a formula for general $n$ propagators.
Subsequently, results for the four-denominator angular integral obtained via differential equations are given in Section \ref{sec: Four denominator angular integral from differential equations}.
Next, we recursively apply the corresponding dimensional shift identity to decompose angular integrals in Section \ref{sec: Scale reduction by dimensional recurrence} into branch integrals, which reduces the number of scales in the master integrals from $n(n+1)/$ to only $n+1$.
We employ this approach in Section \ref{sec: All-order expansion of massless three denominator angular integral by dimensional recurrence} to establish an all-order $\eps$-expansion of the massless three-denominator integral including the resummation of soft singularities, before we conclude in Section \ref{sec: Conclusion}.
Appendix \ref{app: Discussion of specific branch integrals} extends the discussion of branch integrals from Section \ref{sec: Scale reduction by dimensional recurrence} by explicitly going through the recursive construction and $\eps$-expansion for the cases $n\leq 3$.

\section{Euler representation for angular integrals}
\label{sec: Euler representation for angular integral}
Euler integrals are a main focus of the mathematical study of loop amplitudes \cite{Salvatori:2024nva}.
Hence it is natural to also look for such representations for phase-space integrals.
A first step in this direction was taken in \cite{Smirnov:2024pbj}, where such a representation was algorithmically constructed for angular integrals to be used in the study of the small mass asymptotics of angular integrals.
Angular integrals admit the simple and symmetric Euler representation
\begin{align}
\Omega_{j_1,\dots,j_n}(\lbrace v_{kl}\rbrace;\eps)=\frac{2^{2-j-2\eps}\pi^{1-\eps}\,\Gamma(1-\eps)}{\Gamma(2-j-2\eps)\prod_{k=1}^n \Gamma(j_k)}
\int_0^\infty\frac{\prod_{k=1}^n \dx t_k\,t_k^{j_k-1}}{\left(1+\sum_{k=1}^n t_k+\sum_{k\leq l=1}^n \tilde{v}_{kl}\,t_k t_l\right)^{1-\eps}},
\label{eq: simple Euler representation for angular integrals}
\end{align}
which we derive in this section.
The integral representation has striking structural similarity to the Lee-Pomeransky representation for loop integrals \cite{Lee:2013hzt}.
Specifically setting the dimension as $d=2-2\eps$, the loop number $l=1$ and the Lee-Pomeransky polynomial to
\begin{align}
\mathcal{P}(t)=1+\sum_{k=1}^n t_k+\sum_{k\leq l=1}^n \tilde{v}_{kl}\,t_k t_l,
\end{align}
eq.\,\eqref{eq: simple Euler representation for angular integrals} is exactly of the form of a Lee-Pomeransky representation up to overall powers of $2$ and $\pi$.
This representation is considerably simpler than the form given by the authors in \cite{Smirnov:2024pbj} which was used in the investigation of the small mass asymptotics of angular integrals.
In deriving eq.\,\eqref{eq: simple Euler representation for angular integrals}, we start from Somogyi's general Mellin-Barnes representation \cite{Somogyi:2011}
\begin{align}
\Omega_{j_1,\dots,j_n}(\lbrace v_{kl}\rbrace;\eps)=&\frac{2^{2-j-2\eps}\pi^{1-\eps}}{\Gamma(2-j-2\eps)\prod_{k=1}^n \Gamma(j_k)}\int_{-\iu\infty}^{\iu\infty}\left(\prod_{k=1}^n\prod_{l=k}^n\frac{\dx z_{kl}}{2\pi\iu}\,\tilde{v}_{kl}^{z_{kl}}\right)\nonumber\\
&\times\left(\prod_{k=1}^n \Gamma(j_k+z_k)\right)\Gamma(1-j-\eps-z)\,,
\end{align}
where $j=\sum_{k=1}^n j_k$, $z_k=\sum_{l=1}^k z_{lk}+\sum_{l=k}^n z_{kl}$, $z=\sum_{k=1}^n\sum_{l=k}^n z_{kl}$, and normalized scalar products $\tilde{v}_{kl}=v_k\cdot v_l/2$ for $k\neq l$ and $\tilde{v}_{kk}=v_k^2/4$.
Noting that
\begin{align}
\underbrace{\sum_{k=1}^n\sum_{l=k}^n (-z_{kl})}_{=-z}+\underbrace{\sum_{k=1}^n(j_k+z_k)}_{=j+2z}+1-j-\eps-z=1-\eps
\end{align}
we can combine all $z_{kl}$-dependent Gamma functions to a single multi-variable Beta function which can in turn be expressed as an Euler integral
\begin{align}
&\Gamma(1-\eps)\mathrm{B}\left(\lbrace -z_{kl} \rbrace,\lbrace j_k+z_k\rbrace,1-j-\eps-z \right)
=\Gamma(1-\eps)\int_0^\infty \left(\prod_{k=1}^n\prod_{l=k}^n\frac{\dx t_{kl}}{t_{kl}}\right)\left(\prod_{k=1}^n\frac{\dx t_k}{t_k}\right)\frac{\dx t}{t}\,
\nonumber\\
&\hspace{5cm}\times
t_{kl}^{-z_{kl}}\,t_k^{j_k+z_k}\,t^{1-j-\eps-z}\,\delta\!\left(1-\sum_{k\leq l=1}^n t_{kl}-\sum_{k=1}^n t_k-t\right).
\end{align}
The remaining MB integrals can be evaluated in terms of delta functions as
\begin{align}
\int_{-\iu\infty}^{\iu\infty}\prod_{k=1}^n\prod_{l=k}^n\frac{\dx z_{kl}}{2\pi \iu}\left(\frac{\tilde{v}_{kl}\,t_k\,t_l}{t_{kl}\,t}\right)^{z_{kl}}
=\prod_{k=1}^n\prod_{l=k}^n\delta\!\left(1-\frac{\tilde{v}_{kl} \,t_k\,t_l}{t_{kl}\,t}\right)
=\prod_{k=1}^n\prod_{l=k}^n t_{kl}\,\delta\!\left(t_{kl}-\frac{\tilde{v}_{kl} \,t_k\,t_l}{t}\right).
\end{align}
Now substituting $t_k\rightarrow t\,t_k$ and $t_{kl}\rightarrow t\,t_{kl}$ leads us to
\begin{align}
\Omega_{j_1,\dots,j_n}(\lbrace v_{kl}\rbrace;\eps)=&\frac{2^{2-j-2\eps}\pi^{1-\eps}\Gamma(1-\eps)}{\Gamma(2-j-2\eps)\prod_{k=1}^n \Gamma(j_k)}\int_0^\infty\left(\prod_{k=1}^n\prod_{l=k}^n\dx t_{kl}\right)\left(\prod_{k=1}^n\frac{\dx t_k}{t_k}\right)\frac{\dx t}{t}t_k^{j_k}\,t^{1-\eps}\nonumber\\
&\times\prod_{k=1}^n\prod_{l=k}^n\delta(t_{kl}-\tilde{v}_{kl}t_k t_l)\,\delta\left(1-t\left(1+\sum_{k=1}^n t_k+\sum_{k\leq l=1}^n t_{kl}\right)\right).
\end{align}
Finally, evaluating the delta functions in $t_{kl}$ and $t$, we obtain the integral representation in eq.\,\eqref{eq: simple Euler representation for angular integrals}.

\section{IBP relations and reduction to master integrals for angular integrals with $n$ denominators}
\label{sec: IBP relations}
IBP relations for angular integrals can be established in a very similar fashion to how one would do for loop integrals, with the necessary modifications described in \cite{Haug:2024yfi}.
Due to the structural simplicity present in the case of angular integrals, it is possible to symbolically combine the IBP relations to explicit identities that either only raise or lower indices.
Hence, a reduction to master integrals is possible without invoking Laporta's algorithm \cite{Laporta:2000}.
Extending on \cite{Lyubovitskij:2021} and \cite{Haug:2024yfi} where the cases of $n=2$ and $n=3$ denominators have been covered, respectively, we  did the very same exercise for $n=4$ denominators. 
Doing so, one notices a common structure in the recursion relations which we conjecture to hold true for all $n$.

The recursion that lowers the $k$-th index while also lowering the sum $j=\sum_i j_i$, applicable if $j_k\neq 1$, reads
\begin{align}
I_{j_1\dots j_n}=&\frac{1}{X_n(j_k-1)}
\left\lbrace
\left[(j+1-d)X_{4,\bar{k}}+(j_k-1)X_4^{(k,k)}+\sum_{i\neq k} j_i X_n^{(k,i)}
\right]\,\mathbf{\hat{j}^-_k}
\right.\nonumber\\
&\left.
+\sum_{i\neq k}(j_k-1)X_n^{(k,i)}\,\mathbf{\hat{j}^-_i}
+\sum_{i\neq k}(d-j-1)X_n^{(k,i)}\,\mathbf{\hat{j}^-_k}
\mathbf{\hat{j}^-_i}
+\sum_{\underset{i\neq l}{i\neq k,l\neq k}}j_l X_n^{(k,i)}\,\mathbf{\hat{j}^-_k}\,\mathbf{\hat{j}^-_i}\,\mathbf{\hat{j}^+_l}
\right.\nonumber\\
&\left.
+(d-j-1)X^{(1,1)}\left(\mathbf{\hat{j}^-_k}\right)^2
+\sum_{i\neq k} j_i X_n^{(k,k)}\left(\mathbf{\hat{j}^-_k}\right)^2\mathbf{\hat{j}^+_i}
\right\rbrace
I_{j_1\dots j_n}
\label{eq: General index lowering identity}
\end{align} 
where $\mathbf{\hat{j}^\pm_i}$ are the raising (resp. lowering) operators of the $i$-th index, $X_n$ denotes the \textit{Gram determinant}
\begin{align}
X_{n}=(-1)^{n-1}\det\!\left(v_i\cdot v_j\right)_{i,j=1\dots n}\,,
\end{align}
$X_{n,\bar{k}}$ the \textit{Gram-Cramer determinant}, i.e. the Gram determinant whit the $k$-th column replaced by ones
\begin{align}
X_{n,\bar{k}}=(-1)^{n-1}\det\!\left((1-\delta_{jk})\,v_i\cdot v_j+\delta_{jk}\right)_{i,j=1\dots n}\,,
\end{align}
and $X_n^{(k,l)}$ the \textit{Gram cofactors} where the $k$-th row and $l$-th column have been deleted from the Gram determinant
\begin{align}
X_{n}^{(k,l)}=(-1)^{n-1}\det\!\left(v_i\cdot v_j\right)_{i,j=1\dots n, i\neq k, j\neq l}\,.
\end{align}

The corresponding identity for raising the $k$-th index, which also raises the sum of indices $j$, reads
\begin{align}
I_{j_1\dots j_n}=&\frac{\mathbf{\hat{j}^+_k}}{(3+j-d)\,Y_n^{(k,k)}}
\left\lbrace
\left[(3+j-d)Y_n^{(k,k)}+(1+j_k)X_{n,\bar{k}}+\sum_{i\neq k}j_i(1-v_{ki})X_{n,\bar{k}}^{(k,i)}\right]
\right.
\nonumber\\
&\left.
+(1+j_k)\left((1-v_{kk})X_n^{(k,k)}-X_{n,\bar{k}}\right)\mathbf{\hat{j}^+_k}
+\sum_{i\neq k}j_i(1-v_{ki})X_n^{(k,k)}\,\mathbf{\hat{j}^+_i}
\right.
\nonumber\\
&\left.
+\sum_{i\neq k}(1+j_i)(1-v_{kk})X_{n,\bar{k}}^{(k,i)}\,\mathbf{\hat{j}^+_k}\mathbf{\hat{j}^-_i}
+\sum_{\underset{i\neq l}{i\neq k,l\neq k}}j_i(1-v_{ki})X_{n,\bar{k}}^{(k,l)}\,\mathbf{\hat{j}^+_i}\mathbf{\hat{j}^-_l}
\right\rbrace I_{j_1\dots j_n}
\label{eq: General index raising identity}
\end{align}
where also \textit{Gram-Cramer cofactors} $X_{n,\bar{k}}^{(k,l)}$ appear, i.e. Gram-Cramer determinants with the $k$-th row and $l$-th column deleted
\begin{align}
X_{n,\bar{k}}^{(k,l)}=(-1)^{n-1}\det\!\left((1-\delta_{jk})\,v_i\cdot v_j+\delta_{jk}\right)_{i,j=1\dots n,i\neq k, j\neq l}\,.
\end{align}
We also introduced $Y_n$ to denote the \textit{Euclidean Gram determinant}
\begin{align}
Y_{n}=\det\!\left(1-v_i\cdot v_j\right)_{i,j=1\dots n}
\end{align}
with its cofactors 
\begin{align}
Y_{n}^{(k,l)}=\det\!\left(1-v_i\cdot v_j\right)_{i,j=1\dots n, i\neq k, j\neq l}\,.
\end{align}
Remarkably, these recursion relations allow, in principle --- restricted only by memory and computation time --- for an algorithmic reduction of $I_{j_1\dots j_n}$ to master integrals for arbitrary $n$ without using Laporta's algorithm \cite{Laporta:2000}.
To be specific, the reduction can be performed according to the following algorithm, applied to each integral:
\begin{itemize}
\item[1.] Set $n$ to the number of non-zero indices.
\item[2.] \texttt{IF} there is at least one negative index, use eq.\,\eqref{eq: General index raising identity} for $n$ denominators to raise the least negative index. This will eventually reduce all negative indices to zero and lower $n$.
\item[3.] \texttt{ELSE IF} there is an index larger than $1$, use eq.\,\eqref{eq: General index lowering identity} for $n$ denominators to reduce the largest index. This will eventually reduce all positive indices to $1$.
\item[4.] \texttt{ELSE} return the integral as a master integral.
\end{itemize}
This process results in a system with in general $2^n$ master integrals, those with $j_i\in\lbrace 0,1\rbrace$.
In the remainder of this work, we focus on these master integrals starting by discussing their behavior under dimensional shift.

\section{Dimensional shift relation}
\label{sec: Dimensional shift relation}
In \cite{Haug:2024yfi}, a general dimensional recurrence formula that connects angular integrals in $d$ and $d+2$ dimensions with different propagator powers $j_i$ was proven.
Combining this with the IBP reduction of Section \ref{sec: IBP relations} we can generate identities between master integrals in different dimensions.
For four denominators we find the dimensional shift formula
\begin{align}
I_{1111}(d)=\frac{1}{X_{4}}\Big[
&X_{4,\bar{4}} I_{1110}(d)+X_{4,\bar{3}} I_{1101}(d)+X_{4,\bar{2}} I_{1011}(d)+X_{4,\bar{1}} I_{0111}(d)
\nonumber\\
&+\frac{5-d}{d-3}\,Y_{4}\,I_{1111}(d+2)\Big].
\label{eq: dim-shift 4}
\end{align}
In contrast to the case of $n=3$ discussed in \cite{Haug:2024yfi}, here the $d+2$-dimensional part is not suppressed by a power of $\eps$ in $d=4-2\eps$ dimensions.
Consequently, this term contributes at order $\eps^0$.
Generalizing the cases $n=1,2,3,4$, we conjecture the following form for general $n$:
\begin{align}
I_{\iunderbrace{1\dots 1}_n}(d)=\frac{1}{X_{n}}\left[
\sum_{i=1}^n X_{n,\bar{i}}\,I_{1 \underset{\underset{i\text{-th}}{\uparrow}}{\dots 0\dots} 1}(d)+\frac{n+1-d}{d-3}\, Y_{n}\,I_{\iunderbrace{1\dots 1}_n}(d+2)\right].
\label{eq: dim-shift general}
\end{align}
The coefficients $X_n$, $X_{n,\bar{i}}$, and $Y_n$ have been introduced in Section \ref{sec: IBP relations}.

Interestingly, the dimensional shift identity can be viewed as a direct generalization of partial fractioning.
In a case where the Euclidean Gram determinant $Y_n$ vanishes --- which happens if the denominator vectors become spatially linearly dependent --- eq.\,\eqref{eq: dim-shift 4} reduces to a partial fractioning identity between $n$- and $n-1$-denominator integrals in $d$ dimensions.
We note that in a situation where the denominator vectors $v_i$ are confined to the physical four dimensional subspace, a maximum number of three vectors may have linear independent spatial parts, thus always $Y_4=0$ in this case. 
In the following, we do not make this assumption.

By repeated use of eq.\,\eqref{eq: dim-shift general} one can express the angular integral with $2n+1$ denominators in terms of $2n$-denominator integrals up to terms vanishing for $d\rightarrow 4$.
Also, these equations can be used to set up a calculation based on dimensional recurrence \cite{Tarasov:1996br}.
This will be explored in Section \ref{sec: Scale reduction by dimensional recurrence} in detail.
Before, we will however use eq.\,\eqref{eq: dim-shift 4} to proceed analogously to \cite{Haug:2024yfi} to establish results for the four denominator angular integral.

\section{Four denominator angular integral from differential equations}
\label{sec: Four denominator angular integral from differential equations}

\begin{figure}
\centering
\includegraphics[width=0.8\textwidth]{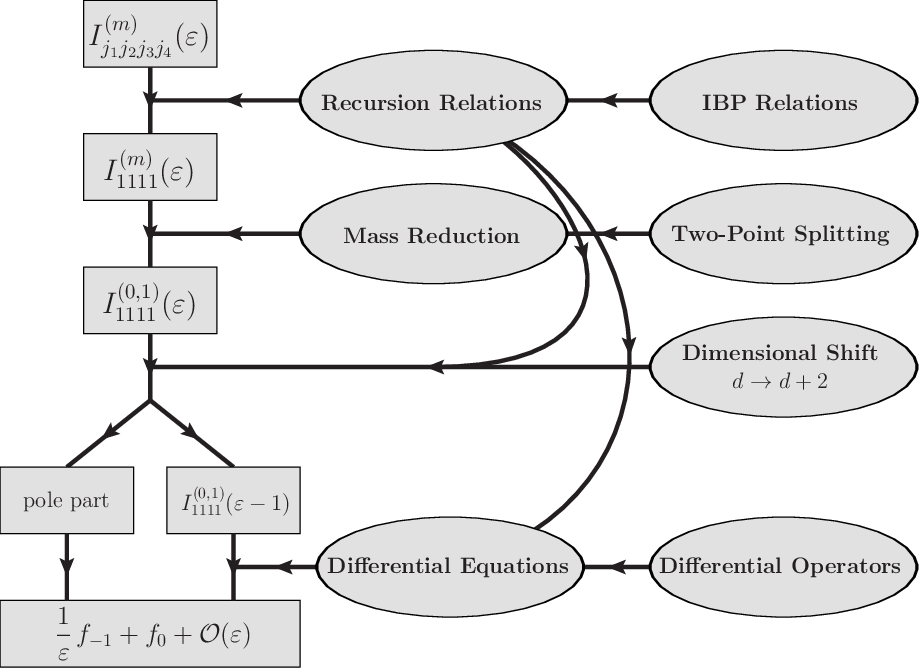}
\caption{This flowchart provides an overview of the calculation of the $\eps$-expansion of the general four-denominator angular integral $I_{j_1j_2 j_3 j_4}^{(m)}$. 
In a first step, recursion relations derived from IBP relations (see sec.\,\ref{sec: IBP relations}) are used for a reduction to the master integral $I_{1111}^{(m)}$. 
In a second step, the double-, triple-, and quadruple-massive integrals are expressed in terms of massless and single-massive ones through mass reduction formulae derived from the two-point splitting lemma (see sec.\,4 of \citep{Haug:2024yfi} for details).
In a third step, a combination of a dimensional shift identity, relating integrals in $d$ and $d+2$ dimensions, with the recursion relations allows for the determination of the pole part and some finite contributions in terms of known three denominator integrals (see sec.\,\ref{sec: Dimensional shift relation}). 
In a final step, the order $\eps^0$ contribution is calculated by applying the method of differential equations --- requiring suitable differential operators for angular integrals (see sec.\,5 of \citep{Haug:2024yfi}) and again making use of the recursion relations --- to the massless and single-massive master integral in $d=6-2\eps$ dimensions.
Graphic created with JaxoDraw \cite{Binosi:2003}.}
\label{fig: Flowchart of calculation}
\end{figure}

Since the method of applying the differential equation technique has been outlined in detail in \cite{Haug:2024yfi} and since going from $3$ to $4$ denominators is a rather straightforward exercise in this case, we only present the results here.
An overview of the involved steps is given in Figure \ref{fig: Flowchart of calculation}.
As mentioned already in the introduction, we want to stress that this approach leads to remarkably compact results --- compare them, e.g., with the one-mass result in \cite{Salvatori:2024nva}.
Calculating $I_{1111}^{(m)}(d+2)$ for $m=0,1$ by differential equations and setting $J_{1111}^{(m),d=6}\equiv\sqrt{X_{4}^{(m)}}\,I_{1111}^{(m),d=6}$, we find
\begin{align}
J_{1111}^{(0),d=6}=2\pi\left[\clausen\!\left(2\theta_{23}^{(0)}\right)+\clausen\!\left(2\theta_{24}^{(0)}\right)+\clausen\!\left(2\theta_{34}^{(0)}\right)\right]
\label{eq: J1111 massless}
\end{align}
where
\begin{align}
\theta_{ij}^{(m)}=\arctan\!\left(\frac{\sqrt{X_4^{(m)}}}{v_{il}v_{jk}+v_{ik}v_{jl}-v_{ij}v_{lk}}\right)\,,
\end{align}
with pairwise distinct indices $i,j,k,l=1,\dots,4$.
For the single-massive case we find
\begin{align}
J_{1111}^{(1),d=6}=\pi
&\left[2\,\clausen\!\left(2\theta_{23}^{(1)}\right)
+2\,\clausen\!\left(2\theta_{24}^{(1)}\right)
+2\,\clausen\!\left(2\theta_{34}^{(1)}\right)
+\clausen\!\left(2\theta_{23}^{(1)}
+2\theta_{23}^{(0)}\right)\right.
\nonumber\\&
+\clausen\!\left(2\theta_{23}^{(1)}-2\theta_{23}^{(0)}\right)
-\clausen\!\left(4\theta_{23}^{(1)}\right)
+\clausen\!\left(2\theta_{24}^{(1)}+2\theta_{24}^{(0)}\right)
+\clausen\!\left(2\theta_{24}^{(1)}-2\theta_{24}^{(0)}\right)
\nonumber\\
&
\left.-\clausen\!\left(4\theta_{24}^{(1)}\right)
+\clausen\!\left(2\theta_{34}^{(1)}+2\theta_{34}^{(0)}\right)+\clausen\!\left(2\theta_{34}^{(1)}-2\theta_{34}^{(0)}\right)-\clausen\!\left(4\theta_{34}^{(1)}\right)\right]
.
\label{eq: J1111 one mass}
\end{align}
Note that in the massless limit, the first three terms of eq.\,\eqref{eq: J1111 one mass} straightforwardly reduce to the massless integral from eq.\,\eqref{eq: J1111 massless}, while the remaining terms cancel.
In comparison to the analogous three denominator result in $d=6$ given in \cite{Haug:2024yfi}, the four-denominator result has a strikingly similar structure, especially in the massless case, but is even more compact.
Also we note that when going from $n=3$ to $n=4$, the Minkowski Gram determinant $X_4$ takes the role of the Euclidean Gram determinant $Y_3$ in the $\theta$ arguments of the Clausen functions.
Details about the latter may be found e.g. in Appendix H of \cite{Haug:2024yfi}. 

By a mass reduction \cite{Lyubovitskij:2021,Haug:2024yfi}, with the help of the two-point splitting lemma, we can construct the double and triple massive integrals from eqs.\,\eqref{eq: J1111 massless} and \eqref{eq: J1111 one mass}.
With the shorthand notation
\begin{align}
J_{1111}^{(m),{d=6}}(v_i,v_j,v_k,v_l)\equiv J^{(m)}(i,j,k,l)\,,
\end{align}
and the auxiliary massless vectors
\begin{align}
v_{(ij)}\equiv (1-\lambda_{(ij)})\,v_i+\lambda_{(ij)}\,v_j\,,
\end{align}
where 
\begin{align}
\lambda_{(ij)}=\frac{v_{ij}-v_{ii}-\sqrt{v_{ij}^2-v_{ii}v_{jj}}}{2v_{ij}-v_{ii}v_{jj}}
\end{align}
we find for the multi-mass integrals
\begin{align}
J^{(2)}(1,2,3,4)&=J^{(0)}((12),(21),3,4)-J^{(1)}(1,(12),3,4)-J^{(1)}(2,(21),3,4)\,,\\
J^{(3)}(1,2,3,4)&=J^{(1)}(1,(23),(32),4)-J^{(2)}(1,2,(23),4)-J^{(2)}(1,3,(32),4)\,,\\
J^{(4)}(1,2,3,4)&=J^{(2)}(1,2,(34),(43))-J^{(3)}(1,2,3,(34))-J^{(3)}(1,2,4,(43))\,.
\end{align}

Combining these with the dimensional shift formula eq.\,\eqref{eq: dim-shift 4} we have for the four denominator integral in $d=4-2\eps$
\begin{align}
I_{1111}^{(m)}(v_1,v_2,v_3,v_4;\eps)=
\frac{1}{X_4}
\left[\vphantom{\frac{XXX}{XXX}}\right.
&X_{4,\bar{4}} I_{111}^{(m)}(v_1,v_2,v_3;\eps)+X_{4,\bar{3}} I_{111}^{(m)}(v_1,v_2,v_4;\eps)
\nonumber\\
&+X_{4,\bar{2}} I_{111}^{(m)}(v_1,v_3,v_4;\eps)+X_{4,\bar{1}} I_{111}^{(m)}(v_2,v_3,v_4;\eps)
\nonumber\\
&\left.+\frac{Y_4}{\sqrt{X_4}}J_{1111}^{(m),d=6}(v_1,v_2,v_3,v_4)\right]+\mathcal{O}(\eps)\,.
\label{eq: result I1111}
\end{align}
The three-denominator integrals appearing in this expression have been reported in \cite{Haug:2024yfi} all other quantities are defined above.

For the case of a single mass, this result is in agreement with \cite{Salvatori:2024nva} and has been additionally checked numerically with \texttt{FIESTA} \citep{FIESTA3:2014,FIESTA4:2016,Smirnov:2021rhf}.
It is worth noting that in contrast to the form reported in \cite{Salvatori:2024nva}, the result \eqref{eq: result I1111} is manifestly real-valued for positive Gram determinant and has transparent symmetry properties with respect to interchange of the vectors $v_i$.

\section{Scale reduction by dimensional recurrence}
\label{sec: Scale reduction by dimensional recurrence}
The angular integral with $n$ denominators depends on $n$ vectors, i.e. on $n(n-1)/2$ scalar products between them, and additionally on $0\leq m\leq n$ masses.
As a rule of thumb: the higher the number of scales, the more complicated the calculation of the integral.
Hence since the very beginning of the study of angular integrals in QCD, and similarly for loop integrals \citep{tHooft:1978,Buza:1995ie}, reducing the number of scales has been of key interest.
The only tool used so far for angular integrals has been partial fractioning decomposition.
In the case of linear dependent vectors this allows to reduce the number of vectors the integral depends on.
A generalization of this extensively employed technique in \citep{Lyubovitskij:2021}, allowing for a reduction of multi-mass integrals to integrals with no more than one mass,
made the calculation of previously inaccessible integrals, like those in Appendix A.3, eqs.\,(159) to (163), of \cite{Blumlein:2020} for general $d=4-2\eps$, possible in \cite{Lyubovitskij:2021,Haug:2024yfi}.
This exemplifies the power of expressing master integrals in terms of integrals with fewer scales.
A powerful technique that achieves this for loop integrals is Tarasov's dimensional recurrence \cite{Tarasov:1996br,Tarasov:2000sf}.
Borrowing from this approach for angular integrals, we will see that iterating the dimensional shift relation eq.\,\eqref{eq: dim-shift general} allows for a reduction from $n(n-1)/2+m$ scales to a sum of terms with only $n$ or $n+1$ scales.
Any of these contributions belongs to one of only two classes depending on whether the `root' vector is massive or massless.
So, overall, the task of calculating $n+1$ integrals with $n(n-1)/2+m$ scales where $m$ takes values from $0$ to $n$ is reduced to the calculation of just two integrals with $n$ and $n+1$ scales each. We will call these the \textit{massless/massive branch integrals} $\mathcal{B}^{(0/1)}_n$, which will be defined below. 

In this section we will iterate the dimensional shift relation of Section \ref{sec: Dimensional shift relation} to a dimensional recurrence formula which we will subsequently use to establish a decomposition formula with the scale reduction properties described above.
From there we will explore the utility of the devised formula to establish the all-order expansion of the massless three denominator integral and several orders for its massive counterpart.
\subsection{From dimensional shift to dimensional recurrence}
We start our exploration from the dimensional shift formula
\begin{align}
I_{\vec{1}_n}(d)=\sum_{i=1}^n x_{n,i} I_{\vec{1}_{n,i}}(d)+\frac{n+1-d}{d-3}\,y_n\, I_{\vec{1}_n}(d+2)\,,
\label{eq: Dim shift abbr}
\end{align}
where we use the compact notation $\vec{1}_n=\underbrace{1,\dots,1}_n$,  $\vec{1}_{n,i}=1 \underset{\underset{i\text{-th}}{\uparrow}}{\dots 0\dots} 1$, $x_{n,i}=X_{n,i}/X_n$, and $y_n= Y_n/X_n$.
Here and in the following we assume that the Gram determinants $X_i$ for $i>1$ are non-zero throughout.
We observe that this identity splits the $n$-denominator master integral into $n$ integrals with one denominator less and a remainder contribution which is the original $n$ denominator integral shifted by two dimensions.
For the three- and four denominator integral this representation was used to facilitate the calculation of the $\eps$-expansion by  using differential equations to determine the dimensionally shifted integral.

The idea of shifting the dimension can be taken a step further by iterating the procedure.
Plugging the right-hand side of eq.\,\eqref{eq: Dim shift abbr} into itself repeatedly an infinite series builds up as
\begin{align}
I_{\vec{1}_n}(d)=&\sum_{i=1}^n x_{n,i} I_{\vec{1}_{n,i}}(d)+\frac{n+1-d}{d-3}\,y_n\sum_{i=1}^n x_{n,i} I_{\vec{1}_{n,i}}(d+2)
\nonumber\\
&+\frac{(n+1-d)(n-1-d)}{(d-3)(d-1)}\,y_n^2\sum_{i=1}^n x_{n,i} 
\, I_{\vec{1}_{n,i}}(d+4)
+\dots
\nonumber\\
&+
\frac{(n+1-d)(n-1-d)\dots(n+1-2k-d)}{(d-3)(d-1)\dots (d+2k-3)}\, y_n^{k}\sum_{i=1}^n x_{n,i} 
\, I_{\vec{1}_{n,i}}(d+2k)+\dots
\nonumber\\
&=\sum_{i=1}^n x_{n,i}\sum_{k=0}^\infty \frac{\pochhammer{\frac{d-n-1}{2}}{k}}{\pochhammer{\frac{d-3}{2}}{k}}\,(-y_n)^k\,I_{\vec{1}_{n,i}}(d+2k).
\label{eq: Dim recurrence solution}
\end{align}
After $k$ iterations there is a remainder term proportional to an angular integral in $d+2k+2$ dimensions.
The convergence of the series is assured by the rapid decrease of the angular measure for $d\rightarrow\infty$.
Eq.\,\eqref{eq: Dim recurrence solution} constitutes a series solution to the dimensional recurrence.
An analogous result for loop integrals was discussed in \cite{Tarasov:2000sf}. 

\subsection{From dimensional recurrence to splitting into branches}
We observe that the representation of eq.\,\eqref{eq: Dim recurrence solution} splits the $n$-denominator integral into $n$ parts each depending on dimensionally shifted $n-1$-denominator integrals.
Now we can perform a second round of iterations and use this very same identity on the $n-1$-denominator integrals.
Doing this $n-1$-times the right hand side is expressed as a sum of $n!$ terms, each written as a nested sum over dimensionally shifted one-denominator integrals.
Each of these terms corresponds to a particular permutation of the $n$ vectors $v_1,\dots,\,v_n$ which are `pinched out` by eq.\,\eqref{eq: Dim recurrence solution} one at a time.

\begin{figure}
\includegraphics[width=\textwidth]{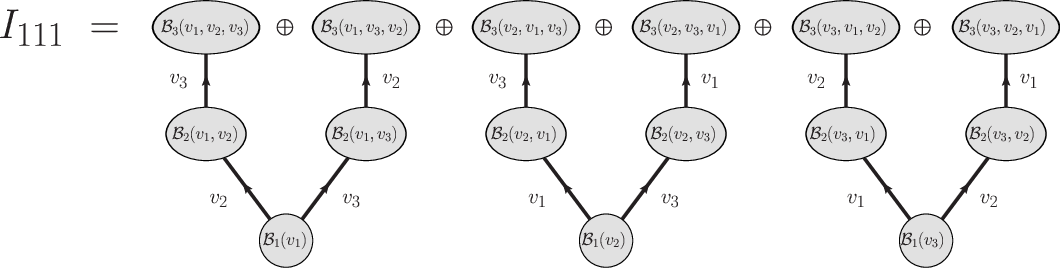}
\caption{Illustration of the decomposition of $I_{111}$ into branches, each associated with a permutation $\sigma\in S_3$. The horizontal $\oplus$-sum over branch integrals $\mathcal{B}_3(v_{\sigma(1)},v_{\sigma(2)},v_{\sigma(3)})$ is performed according to eq.\,\eqref{eq: Branch splitting} with prefactors $x_{2,2}\!\left(v_{\sigma(1)},v_{\sigma(2)}\right)x_{3,3}\!\left(v_{\sigma(1)},v_{\sigma(2)},v_{\sigma(3)}\right)$ inherited from each branch. Vertically growing the branches starting from the `roots' $\mathcal{B}_1(v_i)$ along the arrows by including a new vector $v_j$ is done with eq.\,\eqref{eq: Branch iteration}.
Here, the construction of $\mathcal{B}_{i+1}$ involves summation over $\mathcal{B}_{i}$ in shifted dimensions.}
\label{fig: Branch integrals graphic}
\end{figure}

The result takes the form
\begin{align}
I_{\vec{1}_n}(v_1,\dots,v_n;d=4-2\eps)&=\sum_{\sigma\in S_n}\left(\prod_{i=2}^{n} x_{i,i}\!\left(v_{\sigma(1)},\dots,v_{\sigma(i)}\right)\right)\mathcal{B}_n(v_{\sigma(1)},\dots,v_{\sigma(n)};\eps)
\label{eq: Branch splitting}
\end{align}
with \textit{branch integrals}
\begin{align}
\mathcal{B}_n(v_1,\dots,v_n;\eps)&=\!\!\!\!\!\!\sum_{k_2,\dots,k_n=0}^\infty \left[\prod_{i=2}^n \,c\!\left(\eps-\sum_{j=i+1}^n k_j,i,k_i\right)(-y_i(v_1,\cdots,v_i))^{k_i}\right]I_{1}\!\left(v_1;2-\eps+\sum_{i=2}^n k_i\right)
\end{align}
where the summation coefficients $c(\eps,n,k)$ are given by
\begin{align}
c(\eps,n,k)=\frac{\pochhammer{\frac{3-n}{2}-\eps}{k}}{\pochhammer{\frac{1}{2}-\eps}{k}} \,.
\end{align}
The representation eq.\eqref{eq: Branch splitting} now expresses the $n$-denominator master integral in terms of branch integrals $\mathcal{B}_n$ which are given by (hypergeometric) nested sums over a 'root' one-denominator integral.
While the original integral depends on $n(n-1)/2+m$ scales, these branch integrals depend only on the $n$ Gram-type variables $y_i$ and, if $v_1$ is massive, additionally on $v_{11}$.
Furthermore, while there are $n+1$ different massless/massive configurations for $I_{\vec{1}_n}$, for the $\mathcal{B}_n$ there is only a distinction to be made between massless/massive for the root vector hence there are only two different functions to be calculated for a certain $n$, $\mathcal{B}_n^{(0)}$ and $\mathcal{B}_n^{(1)}$. 
Iteratively, the branch integrals can be calculated as
\begin{align}
\mathcal{B}_{n+1}(v_1,\dots,v_{n+1};\eps)=\sum_{k_{n+1}=0}^\infty
&c(\eps,n+1,k_{n+1})(-y_{n+1}(v_1,\dots,v_{n+1}))^{k_{n+1}}\nonumber\\
&\times\mathcal{B}_{n}(v_1,\dots,v_n;\eps-k_{n+1}).
\label{eq: Branch iteration}
\end{align}
Quite interestingly we note that the set of dimensional shift relations up to $n$ determines the angular integral with $n$ denominators up to the boundary condition at $n=1$. For the massive case we could iterate once more and shift the boundary to $n=0$, thereby reaching a scaleless boundary condition as in the massless case.
It is especially remarkable that the mass-reduction formula used to reduce scales is not required as an additional input but the reduction to only one `relevant' mass is implicitly built in.
Figure \ref{fig: Branch integrals graphic} exemplifies the decomposition into branches for the three-denominator integral $I_{111}$.

In Appendix \ref{app: Discussion of specific branch integrals} we will look at the cases of $n$ up to $3$ to showcase how the sums can be iteratively build up, summed, and turned into $\eps$-expansions.
For the massless $\mathcal{B}_3^{(0)}$ branch integral this works so smoothly that it is even possible to construct a closed form expression valid to all orders in $\eps$ which we will use in the next section.

\section{All-order expansion of massless three denominator angular integral from branch integrals}
\label{sec: All-order expansion of massless three denominator angular integral by dimensional recurrence}
The three-denominator angular integral has recently been calculated by differential equations \cite{Haug:2024yfi}, as well as from Mellin-Barnes integrals \cite{Ahmed:2024pxr}.
In both cases the first terms of the $\eps$-expansion were provided.
Here we take a fresh approach to this integral using the branch integrals constructed from dimensional recurrence in the previous section.

We start from the representation
\begin{align}
I_{111}^{(0)}=\sum_{\sigma\in S_3}\,x_{2,2}\!\left(v_{\sigma(1)},v_{\sigma(2)}\right)\,x_{3,3}\!\left(v_{\sigma(1)},v_{\sigma(2)},v_{\sigma(3)}\right)\mathcal{B}_3^{(0)}\!\left(v_{\sigma(1)},v_{\sigma(2)},v_{\sigma(3)};\eps\right)\,.
\end{align}
Now we can use the $\eps$-expansion of the branch integrals from Appendix \ref{app: Discussion of specific branch integrals}.
Two of the branch integrals always give identical contributions to $I_{111}^{(0)}$, and
we find
\begin{align}
I_{111}^{(0)}(\eps)&=2\pi\sum_{i=1}^3\left\lbrace
\frac{x_{3,i}}{w_i}\left[
-\frac{1}{\eps}+\sum_{N=0}^\infty\eps^N\mathrm{Li}_{N+1}\!\left(1-\frac{2}{w_i}
\right)
\right]
\right\rbrace\nonumber\\
&+
\frac{4\pi\sqrt{Y_3}}{X_3}\,\frac{4^\eps\,\Gamma(1-2\eps)}{\Gamma^2(1-\eps)}\sum_{i=1}^3\sum_{N=0}^\infty\eps^{N+1}
\nonumber\\
&\times
\Big[
(-1)^{N+1}\sum_{\lbrace\pm\rbrace_{N+1}}\mathrm{Im}\,\goncharov(\iu\delta_i,\pm\beta,\pm 1,\dots,\pm 1;1)
\nonumber\\
&\qquad
+\left(\polylog_{N+1}(-y_3)-\polylog_{N+1}\!\left(1-\frac{2}{w_i}\right)\right)\arctan\frac{1}{\delta_i}
\nonumber\\
&\qquad
-\sum_{k=1}^N\polylog_{N-k+1}\!\left(1-\frac{2}{w_i}\right)(-1)^k\sum_{\lbrace\pm\rbrace_{k}}\,\mathrm{Im}\,\goncharov(\iu\delta_i,\pm 1,\dots,\pm 1;1)
\Big],
\label{eq: I111 dim 6 all order result}
\end{align}
with $w_1=v_{23}$, $w_2=v_{13}$, and $w_3=v_{12}$, as well as
\begin{align}
\delta_i=\frac{-w_i+w_j+w_k}{\sqrt{Y_3}}\quad (i,j,k \text{ pairwise different)}\quad \text{and}\quad
\beta=\sqrt{\frac{1+y_3}{y_3}}\,.
\end{align}
The sums over $\lbrace \pm\rbrace_n$ indicate summation over all possible combinations of the $n$ signs appearing in the weights. 
Eq.\,\eqref{eq: I111 dim 6 all order result} is an all-order $\eps$-expansion of the massless three-denominator angular integral in terms of multiple polylogarithms.

For numerical evaluation of the Goncharov polylogarithms \cite{Goncharov:2001}, especially for high orders in $N$ beyond the capabilities of \texttt{PolyLogTools} \cite{Duhr:2019}\footnote{Of course, it is highly unlikely that such a case will occur in practice.}, there are the following one-fold integral representations with elementary integration kernels
\begin{align}
\sum_{\lbrace\pm\rbrace_{N+1}}\mathrm{Im}\,\goncharov(\iu\delta_i,\pm\beta,\pm 1,\dots,\pm 1;1)
=\int_0^1\frac{\dx u}{\beta^2-u}\left[\arctan\frac{\sqrt{u}}{\delta_i}-\arctan\frac{1}{\delta_i}\right]\frac{\log^N(1-u)}{N!}
\end{align}
and
\begin{align}
\sum_{\lbrace\pm\rbrace_{k}}\mathrm{Im}\,\goncharov(\iu\delta_i,\pm 1,\dots,\pm 1;1)
=\int_0^1\dx u\frac{\delta_i}{\delta_i^2+u^2}\frac{\log^k(1-u^2)}{k!}.
\end{align}

Note that $\mathrm{Im}\,\goncharov(i\delta_i,\dots)$ is supposed to always mean
\begin{align}
\mathrm{Im}\,\goncharov(\iu\delta_i,\dots)=\frac{1}{2\iu}\left[\goncharov(\iu\delta_i,\dots)-\goncharov(-\iu\delta_i,\dots)\right].
\end{align}
For real phase-space kinematics, the right-hand side is the correct imaginary part since the other letters are all real-valued and the integration is away from branch cuts.

Comparing the $N=0$ term to the result in \cite{Wunder:2024}, we find numerical agreement.
However, here we have a representation with six weight-two functions $\mathrm{Im}\,\goncharov(\iu\delta_i,\pm\beta;1)$ with a total of four distinct arguments plus a product term with an arcus tangent while the result in \cite{Wunder:2024} is the sum of seven Clausen functions of weight two and is free of product terms.

Having an all-order result at hand, we are now in a position to briefly discuss the behavior of the integral in the soft limit $w_i\rightarrow 0$. 
As is well-known for the two-denominator angular integrals, the massless angular integral may contain additional soft singularities.
Those need to be taken into account whenever performing an additional integration including the soft limit.
A recent physical example can be found in the calculation of the SIDIS phase-space integrals with the radial-angular decomposition method \cite{Ahmed:2024owh}.
For illustrative purpose and since they appear as part of the three-denominator integral we briefly discuss the soft singularities for the much simpler two-denominator massless integral before performing an analogous procedure for the full three-denominator case.
The massless two denominator integral has the all-order expansion\footnote{Setting $w\rightarrow z/(z+z^\prime- z z^\prime)$ and $\eps\rightarrow -\eps/2$, this is the all-order version of eq. (4.22) in \cite{Ahmed:2024owh}.}
\begin{align}
I_{11}^{(0)}(w;\eps)=\frac{2\pi}{w}\,\left[-\frac{1}{\eps}+\sum_{N=0}^\infty \eps^N\,\polylog_{N+1}\left(1-\frac{2}{w}\right) \right].
\end{align}
Besides the overall $1/w$, in every order in $\eps$ this expression has additional logarithmic singularities for $w\rightarrow 0$ which would be an issue when considering an integral of the form
\begin{align}
\int_0^{W}\dx w\, w^{-\eps} I_{11}^{(0)}(w),
\end{align}
where we would like to use the distributional identity
\begin{align}
w^{-1-n\eps}=-\frac{1}{n\eps}\delta(w)+\sum_{N=0}^\infty\frac{(-n \eps)^N}{N!}\left[\frac{\log^N w}{w}\right]_+
\end{align}
on the integrand.
However, we can resum all problematic logarithms using the identity
\begin{align}
-\frac{1}{\eps}+\sum_{N=0}^\infty \eps^N \polylog_{N+1}(x)=(1-x)^\eps \left[-\frac{1}{\eps}+\sum_{N=1}^\infty \eps^N \sum_{m=1}^N(-1)^{m}\mathrm{S}_{N-m+1,m}\left(\frac{x}{x-1}\right)\right]
\label{eq: Polylog resummation}
\end{align}
resulting in the soft-regularized $\eps$-expansion of the massless integral \cite{Lyubovitskij:2021}
\begin{align}
I_{11}^{(0)}(w;\eps)=\frac{2\pi}{w}\left(\frac{w}{2}\right)^{-\eps}\left[-\frac{1}{\eps}+\sum_{N=1}^\infty \eps^N \sum_{m=1}^N(-1)^{m}\mathrm{S}_{N-m+1,m}\left(1-\frac{w}{2}\right)\right].
\end{align}
The $w^{-1-\eps}$ structure of the massless two-denominator angular integral has been known since the early days of QCD \cite{Ellis:1980} and has been a key ingredient to a lot of the phenomenological studies mentioned in the introduction.
In the following we recast the three-denominator integral in an analogous way to allow for similar phenomenological usage.

From our all-order result eq.\,\eqref{eq: I111 dim 6 all order result}, we can identify all logarithms that become large in the limits $w_i\rightarrow 0$.
For real-valued time- or lightlike vectors, the sign of the Minkowski and Euclidean Gram determinants are both positive, hence $0\leq y_3\leq \infty$ and thus $1\leq\beta\leq\infty$.
For the $\delta_i$ there are no restrictions, $-\infty\leq\delta_i\leq\infty$.
However, since they only appear in functions that are bounded for real-valued $\delta$, they do not produce singular logarithms.
The only source of such logarithms are the polylogarithms $\polylog_{n}(-z)$ for $z\rightarrow \infty$ and $\polylog_n(1-2/w_i)$ for $w_i\rightarrow 0$.
These can be resummed using the very same identity as for the two-denominator integral.

Applying eq.\,\eqref{eq: Polylog resummation} to the all-order result eq.\,\eqref{eq: I111 dim 6 all order result} gives, after some rearrangement of the series,
\begin{align}
I_{111}^{(0)}(\eps)&=2\pi\,\sum_{i=1}^3
\frac{x_{3,i}}{w_i}\left(\frac{w_i}{2}\right)^{-\eps}\left[-\frac{1}{\eps}+\sum_{N=1}^\infty \eps^N \mathcal{S}_{N+1}\!\left(1-\frac{w_i}{2}\right) \right]
\nonumber\\
&+4\pi\frac{\sqrt{Y_3}}{X_3}
\,\frac{4^\eps \Gamma(1-2\eps)}{\Gamma^2(1-\eps)}\sum_{i=1}^3
\left\lbrace
\sum_{N=1}^\infty \eps^N\sum_{\lbrace\pm\rbrace_{N}}\Big[\mathcal{G}_{N+1}(\delta_i,\beta)-\mathcal{G}_{N+1}(\delta_i,1)
\Big]
\right.
\nonumber\\
&
\qquad+(1+y_3)^\eps\arctan\frac{1}{\delta_i}
\left[-1+\sum_{N=2}^\infty \eps^N \mathcal{S}_N\!\left(\frac{y_3}{y_3+1}\right)\right]
+\left(\frac{w_i}{2}\right)^{-\eps}
\left[{\arctan\frac{1}{\delta_i}}
\right.
\nonumber\\
&
\qquad\qquad
\left.\left.
+\eps\,\mathcal{G}_2(\delta_i,1)
+
\sum_{N=2}^\infty \eps^N \left(\mathcal{G}_{N+1}(\delta_i,1)-\sum_{k=0}^{N-2}\mathcal{G}_{k+1}(\delta_i,1)\mathcal{S}_{N-k}\!\left(1-\frac{w_i}{2}\right)\right)\right]
\right\rbrace,
\end{align}
with abbreviations
\begin{align}
\mathcal{S}_N(z)&\equiv\sum_{m=1}^{N-1}(-1)^m \mathrm{S}_{N-m,m}(z)
\,,\nonumber\\
\mathcal{G}_{N+1}(\delta_i,\beta)&\equiv(-1)^N\sum_{\lbrace\pm\rbrace_N}\mathrm{Im}\,\goncharov(\iu\delta_i,\pm \beta,\underbrace{\pm 1,\dots,\pm 1}_{N-1};1)\,.
\end{align}
In this final result for the all-order expansion of the massless three denominator integral, the logarithms that can become singular in the soft limit are resummed explicitly.
We note that
\begin{align}
\sum_{i=1}^3\arctan\frac{1}{\delta_i}=-\arctan\frac{\sqrt{Y_3}}{4-w_1-w_2-w_3}\,,
\end{align}
which appeared as an argument in the results of \citep{Haug:2024yfi}.

\section{Conclusion}
\label{sec: Conclusion}
Considerably extending the existing results in the literature, we have investigated a plethora of aspects of multi-propagator angular integrals, transferring methods from loop integration to a phase-space setting.
With a focus on the structure of this particular class of integrals for various numbers of denominators, we uncovered several new features.
The most important ones are
\begin{itemize}
\item a Lee-Pomeransky-like integral representation that very closely resembles a proper Feynman integral,
\item IBP relations cast into recursion relations allowing for a reduction to master integrals without Laporta's algorithm,
\item scale reduction via Tarasov's dimensional recurrence and decomposition into branch integrals.  
\end{itemize}  
Furthermore, we established novel explicit results for the $\eps$-expansion of the four denominator angular integral with an arbitrary number of masses, using a differential equation approach.
Additionally, we presented an all-order $\eps$-expansion for the massless angular integral with three denominators, allowing resummation of soft logarithms.
We hope that these findings are of interest on a theoretical level --- given their strong connection to the study of Feynman integrals --- and are also of value for phenomenology when extending the use of angular integrals to higher-multiplicity processes in the future.

\acknowledgments
F.W. and J.H. are grateful to Werner Vogelsang for his encouragement and the possibility to pursue this work as part of their PhD projects.
We want to thank Stefano Laporta for making us aware of his work on a related class of angular integrals in \citep{Laporta:1994mb}.
The work of V.S. was conducted under the state assignment of Lomonosov Moscow State University
(the derivation of an integral representation and an implementation of dimensional shift relations). It was supported by the Moscow Center for Fundamental and Applied Mathematics of 
Lomonosov Moscow State University under Agreement No. 075-15-2025-345
(deriving a series of new analytic results for angular integrals with four denominators). 
 
\appendix
\section{Discussion of specific branch integrals} 
\label{app: Discussion of specific branch integrals}
We have seen in Section \ref{sec: Scale reduction by dimensional recurrence} that angular integrals can be constructed from the branch integrals introduced there.
In this appendix we will iteratively build up the first few branch integrals $\mathcal{B}_n$ starting from $n=1$.
Doing so, we will see that for the branch integrals there are two distinct representations of interest.
For one of them, one is interested in the $\eps$-expansion of $\mathcal{B}_n$ to express the $\eps$-expansion of $I_{\vec{1}_n}$.
For the other, it is useful to have a representation with an explicitly simple dependence on the dimensionality, since this will be summed over when calculating $\mathcal{B}_{n+1}$.
\subsection{Branch integrals for $n=1$}
For $n=1$, the branch integral is directly equal to the one-denominator angular integral,
\begin{align}
\mathcal{B}_1^{(0/1)}(v_1;\eps)=I_1^{(0/1)}(v_{11};\eps).
\label{eq: Massless root}
\end{align}
Hence it is
\begin{align}
\mathcal{B}_1^{(0)}(v_1;\eps)=-\frac{\pi}{\eps}\,
\end{align}
which has both a simple form for iteration and for use as part of an $\eps$-expansion.
For the massive integral we have the representation
\begin{align}
\mathcal{B}_1^{(1)}(v_1;\eps)=\frac{2\pi}{\sqrt{1-v_{11}}}\left(\frac{v_{11}}{1-v_{11}}\right)^{-\eps}\int_0^{\sqrt{1-v_{11}}}\frac{\dx t}{1-t^2}\left(\frac{t^2}{1-t^2}\right)^{-\eps}
\label{eq: Massive root}
\end{align}
which has a particular simple dependence on $\eps$.
Since $\mathcal{B}_1^{(1)}=I_1^{(1)}$ an expansion in $\eps$ of this term can be found in \citep{Lyubovitskij:2021} and \citep{Wunder:2024}.
The representations eqs.\,\eqref{eq: Massless root} and \eqref{eq: Massive root} will be the root integrals for constructing $B_n^{(0)}$ and $B_n^{(1)}$, respectively.
\subsection{Branch integrals for $n=2$}
Going from $n=1$ to $n=2$ is particularly simple, since $c(\eps,2,k)=1$.
Hence it is
\begin{align}
\mathcal{B}_2^{(0/1)}(v_1,v_2;\eps)=\sum_{k_2=0}^\infty (-y_2(v_1,v_2))^{k_2}
\mathcal{B}_1^{(0/1)}(v_1;\eps-k_2)\,.
\end{align}
Thus, for the massless integral, it is
\begin{align}
\mathcal{B}_2^{(0)}(v_1,v_2;\eps)=\pi \sum_{k_2=0}^\infty \frac{\left(-y_2(v_1,v_2)\right)^{k_2}}{k_2-\eps}
\end{align}
and for the massive integral we find by summing a geometric series
\begin{align}
\mathcal{B}_2^{(1)}(v_1,v_2;\eps)=\frac{2\pi}{\sqrt{1-v_{11}}}
\left(\frac{v_{11}}{1-v_{11}}\right)^{-\eps}\int_0^{\sqrt{1-v_{11}}}\frac{\dx t}{1-(1-\alpha_2(v_1,v_2))t^2}\left(\frac{t^2}{1-t^2}\right)^{-\eps}
\end{align}
with $\alpha_2(v_1,v_2)=v_{11} y_2(v_1,v_2)/(1-v_{11})$.
Both the massless and massive integral are in a form well suited to a further summation going to $n=3$ due to the rather simple dependence on $\eps$.
\subsection{Branch integrals for $n=3$}
Going from $n=2$ to $n=3$, there is an actual Pochhammer coefficient in the $k_3$ sum.
It is
\begin{align}
\mathcal{B}_3^{(0/1)}(v_1,v_2,v_3;\eps)=\sum_{k_3=0}^\infty \frac{\pochhammer{-\eps}{k_3}}{\pochhammer{\frac{1}{2}-\eps}{k_3}}\,(-y_3(v_1,v_2,v_3))^{k_3} \mathcal{B}_2^{(0/1)}(v_1,v_2;\eps-k_3)\,,
\end{align}
so the $k_3$ sum cannot be carried out as easily as before.
In the following, we will focus on extracting the $\eps$-expansion for $\mathcal{B}_3$.
For this, it makes sense to isolate the $k_3=0$ term since the remaining sum will have a common global factor of $\eps$.
Furthermore, and even more crucially, it turns out that this is very helpful for performing an expansion in terms of uniform transcendental weight.\footnote{This step is analogous to the dimensional shift from $d$ to $d+2$ used in \cite{Haug:2024yfi} to calculate the three denominator integral.}
So, suppressing the $v_i$ variables for compactness, we obtain
\begin{align}
\mathcal{B}_3^{(0/1)}(\eps)=\mathcal{B}_2^{(0/1)}(\eps)+\frac{2\eps y_3}{1-2\eps}\sum_{k_3=0}^\infty \frac{\pochhammer{1-\eps}{k_3}}{\pochhammer{\frac{3}{2}-\eps}{k_3}}\,(-y_3)^{k_3} \mathcal{B}_2^{(0/1)}(\eps+1-k_3).
\end{align}
To turn this hypergeometric series representation into an $\eps$-expansion, we make use of the integral representation
\begin{align}
\frac{\pochhammer{1-\eps}{k}}{\pochhammer{\frac{3}{2}-\eps}{k}}=\frac{4^\eps\,\Gamma(2-2\eps)}{\Gamma^2(1-\eps)}\int_0^1\dx u\,(1-u^2)^{k-\eps}\,.
\end{align}
It is quadratic in the integration variable but avoids square roots that would appear in more standard representations of the Beta function.
Plugging this in we have
\begin{align}
\mathcal{B}_3^{(0/1)}(\eps)=\mathcal{B}_2^{(0/1)}(\eps)+&\frac{2\eps y_3}{1-2\eps}\frac{4^\eps\,\Gamma(2-2\eps)}{\Gamma^2(1-\eps)}\int_0^1\dx u\,(1-u^2)^{-\eps}\nonumber\\
&\times\sum_{k_3=0}^\infty \,(-(1-u^2)y_3)^{k_3} \mathcal{B}_2^{(0/1)}(\eps+1-k_3).
\end{align}
For summing the $k_3$ series it makes a difference whether we deal with $\mathcal{B}_2^{(0)}$ or $\mathcal{B}_2^{(1)}$.
\subsubsection{Massless root integral}
Here we find
\begin{align}
&\sum_{k_3=0}^\infty \left(-y_3 (1-u^2)\right)^{k_3} \mathcal{B}_2^{(0)}(\eps-k_3-1)
\nonumber\\
&=\frac{\pi}{y_3}\,\frac{1}{u^2+\left(\frac{y_2}{ y_3}-1\right)}\,\sum_{k=0}^\infty \eps^k\left[\polylog_{k+1}\left(-y_3(1-u^2)\right)-\polylog_{k+1}\left(-y_2\right)\right]\,.
\end{align}
It remains to calculate the $u$ integral.
The $u$-denominator can be written as
\begin{align}
\frac{1}{u^2+\left(\frac{y_2}{y_3}-1\right)}
=\frac{1}{\delta_3}\frac{\delta_3}{u^2+\delta_3^2}
\end{align}
with
\begin{align}
\delta_3=\sqrt{\frac{y_2}{y_3}-1}\,.
\end{align}
Hence, with the notation cleaned up, we need to calculate
\begin{align}
\mathcal{B}_3^{(0)}(\eps)
&=\mathcal{B}_2^{(0)}(\eps)
+\frac{\pi}{\delta_3}\,\frac{2\eps}{1-2\eps}\,\frac{4^\eps \Gamma(2-2\eps)}{\Gamma^2(1-\eps)}\sum_{k=0}^\infty \eps^k\nonumber\\
&\times
\int_0^1\dx u\,\frac{\delta_3\,(1-u^2)^{-\eps}}{u^2+\delta_3^2}
\left[\polylog_{k+1}\left(-y_3(1-u^2)\right)-\polylog_{k+1}\left(1-\frac{2}{v_{12}}\right)\right].
\label{eq: B30 intermediated step 1}
\end{align}
To evaluate the parametric $u$-integral, we convert the classical polylogarithms with argument $-y_3(1-u^2)$ into Goncharov polylogarithms with argument $u$ \cite{Goncharov:2001, Duhr:2019}.
This can be done in an algorithmic way by taking derivatives with respect to $u$ and re-integrating, resulting in
\begin{align}
\polylog_{k+1}(-y_3(1-u^2))=\sum_{j=0}^k \polylog_{k+1-j}(-y_3)\,\frac{1}{j!}\log^j(1-u^2)-\sum_{\lbrace\pm\rbrace_{k+1}}\goncharov(\underbrace{\pm 1,\dots,\pm 1}_k,\pm\beta;u)
\end{align}
with
\begin{align}
\beta=\sqrt{\frac{X_3+Y_3}{Y_3}}=\sqrt{\frac{-\lambda(v_{12},v_{13},v_{23})}{Y_3}}=\sqrt{1+\frac{1}{y_3}}
\end{align}
where $\lambda(x,y,z)=x^2+y^2+z^2-x y-x y - y z$ is the Källen function and the sum runs over all permutations of signs.

Reorganizing the summation in eq.\,\eqref{eq: B30 intermediated step 1} as
\begin{align}
\sum_{k=0}^\infty \eps^k (1-u)^{-\eps} \varphi(k)=\sum_{N=0}^\infty \eps^N \sum_{k=0}^N \frac{(-1)^{N-k}}{(N-k)!}\log^{N-k}(1-u^2) \varphi(k),
\end{align}
writing the logarithm in Goncharov form according to
\begin{align}
\frac{\log^m(1-u^2)}{m!}=\sum_{\lbrace\pm\rbrace_m}\goncharov(\underbrace{\pm 1,\dots, \pm 1}_{m};u)
\end{align}
and using the identities
\begin{align}
&\sum_{k=0}^N\sum_{\lbrace\pm\rbrace_{k+1}}\frac{(-1)^{N-k}\log^{N-k}(1-u^2)}{(N-k)!}\goncharov(\underbrace{\pm 1,\dots,\pm 1}_{k},\pm\beta;u)
\nonumber\\
&=(-1)^N \sum_{\lbrace\pm\rbrace_{N+1}}\goncharov(\pm\beta,\pm 1,\dots,\pm 1;u)\,
\end{align}
and
\begin{align}
\sum_{k=0}^N\frac{(-1)^{N-k}}{(N-k)!}\log^{N-k}(1-u^2)\sum_{j=0}^l \polylog_{k+1-j}(-z)\frac{1}{j!}\log^j(1-u^2)=\polylog_{N+1}(-y_3)
\end{align}
we obtain
\begin{align}
\mathcal{B}_3^{(0)}(\eps)
&=\mathcal{B}_2^{(0)}(\eps)
+\frac{\pi}{\delta_3}\frac{2\eps}{1-2\eps}\,\frac{4^\eps\,\Gamma(2-2\eps)}{\Gamma^2(1-\eps)}\sum_{N=0}^\infty\eps^N \int_0^1\dx u\,\frac{\delta_3}{u^2+\delta_3^2}
\nonumber\\
&\times
\Big[
(-1)^{N+1}\sum_{\lbrace\pm\rbrace_{N+1}}\goncharov(\pm\beta,\pm 1,\dots,\pm 1;u)
+\left(\polylog_{N+1}(-y_3)-\polylog_{N+1}\!\left(-y_2\right)\right)
\nonumber\\
&\qquad
-\sum_{k=1}^N\polylog_{N-k+1}\!\left(-y_2\right)\sum_{\lbrace\pm\rbrace_{k}}(-1)^k\goncharov(\pm 1,\dots,\pm 1;u)
\Big].
\end{align}
In this form, evaluating the $u$ integral is now straightforward.
With
\begin{align}
\frac{\delta_3}{u^2+\delta_3^2}=\frac{1}{2 \iu}\left(\frac{1}{u-\iu \delta_3}-\frac{1}{u+\iu \delta_3}\right)=\mathrm{Im}\left[\frac{1}{u-\iu \delta_3}\right]
\end{align}
we then find
\begin{align}
\mathcal{B}_3^{(0)}(\eps)
&=\mathcal{B}_2^{(0)}(\eps)
+\frac{2\pi\eps}{\delta_3}\,\frac{4^\eps\,\Gamma(1-2\eps)}{\Gamma^2(1-\eps)}\sum_{N=0}^\infty\eps^N
\nonumber\\
&\times
\Big[
(-1)^{N+1}\sum_{\lbrace\pm\rbrace_{N+1}}\mathrm{Im}\,\goncharov(\iu\delta_3,\pm\beta,\pm 1,\dots,\pm 1;1)
\nonumber\\
&\qquad
+\left(\polylog_{N+1}(-y_3)-\polylog_{N+1}\!\left(-y_2\right)\right)\arctan\frac{1}{\delta_3}
\nonumber\\
&\qquad
-\sum_{k=1}^N\polylog_{N-k+1}\!\left(-y_2\right)(-1)^k\sum_{\lbrace\pm\rbrace_{k}}\,\mathrm{Im}\,\goncharov(\iu\delta_3,\pm 1,\dots,\pm 1;1)
\Big].
\label{eq: B30 all order result}
\end{align}
This is an all-order $\eps$-expansion of the massless three-denominator branch integral in terms of multiple polylogarithms that was used in Section \ref{sec: All-order expansion of massless three denominator angular integral by dimensional recurrence} of the main text.
We notice that when combining with the branch integral with the prefactors in the full angular integral we can use the identity
\begin{align}
\delta_3^2=\frac{X_{3,3}^2\,X_{2,2}^2}{Y_3\,X_2^2}
\end{align}
which holds whenever $v_{11}=0$.
Dependence on absolute values can be eliminated by dropping them in the prefactor as well as in the weights, replacing
\begin{align}
\delta_3\rightarrow\tilde{\delta}_3=\frac{x_{2,2}X_{3,3}}{\sqrt{Y_3}}\,.
\end{align}

\subsubsection{Massive root integral}
We now turn to the calculation of the $\eps$-expansion of the massive branch integral $\mathcal{B}_3^{(1)}$.
Starting from
\begin{align}
\mathcal{B}_3^{(1)}(\eps)=\mathcal{B}_2^{(1)}(\eps)+\frac{2\eps}{1-2\eps}\,y_3\,\sum_{k_3=0}^\infty\frac{\pochhammer{1-\eps}{k_3}}{\pochhammer{\frac{3}{2}-\eps}{k_3}}(-y_3)^{k_3}\mathcal{B}_1^{(1)}(\eps-1-k_3)
\end{align}
we plug in the integral representation for the Pochhammer symbols and $\mathcal{B}_1^{(1)}$.
The resulting $k_3$ sum is a simple geometric series.
Evaluating this we arrive at
\begin{align}
\mathcal{B}_3^{(1)}(\eps)&=\mathcal{B}_2^{(1)}(\eps)+\frac{2\eps}{1-2\eps}\,\,\frac{4^\eps\Gamma(2-2\eps)}{\Gamma^2(1-\eps)}\,\frac{2\pi}{\sqrt{1-v_{11}}}\left(\frac{v_{11}}{1-v_{11}}\right)^{-\eps}
\nonumber\\
&\times\int_0^{\sqrt{1-v_{11}}}\frac{\dx t}{(1-\alpha_2)t^2-1}
\left(\frac{t^2}{1-t^2}\right)^{-\eps}\int_0^1\dx u \frac{(1-u^2)^{-\eps}}{u^2-\beta^2(t)}
\end{align}
with 
\begin{align}
\beta(t)&=\frac{\sqrt{1-(1-\alpha_{3})t^2}}{t\sqrt{\alpha_3}}\,\quad\text{and}\quad
\alpha_3=\frac{v_{11} y_3}{1-v_{11}}\,.
\end{align}
Calculating the nested integral in terms of generalized polylogarithms requires rationalization of the square root $\beta(t)$ \cite{Besier:2018}.
Substituting $t$ for $s$ given by
\begin{align}
t=\frac{s(2-s)}{\sqrt{1-\alpha_3}\,(2-2s+s^2)}
\end{align}
leads to the integral representation 

\begin{align}
\mathcal{B}_3^{(1)}(\eps)&=\mathcal{B}_2^{(1)}(\eps)+\frac{2\eps}{1-2\eps}\,\,\frac{4^\eps\Gamma(2-2\eps)}{\Gamma^2(1-\eps)}\,\frac{\pi\sqrt{\alpha_3}(1-\alpha_3)^\eps}{\sqrt{1-v_{11}}\sqrt{1-\alpha_2}\sqrt{\alpha_2-\alpha_3}}\left(\frac{v_{11}}{1-v_{11}}\right)^{-\eps}
\nonumber\\
&\qquad\times\int_0^{s_\text{max}}\dx s\,\mathrm{Im}\left[\frac{1}{s-(1+\iu r_+)}-\frac{1}{s-(1+\iu r_-)}\right]
\nonumber\\
&\qquad\qquad\times\left(\frac{s^2 \left(1-\frac{s}{2}\right)^2}{\left(1-\frac{s}{s_{++}}\right)\left(1-\frac{s}{s_{+-}}\right)\left(1-\frac{s}{s_{-+}}\right)\left(1-\frac{s}{s_{--}}\right)}\right) ^{-\eps}\nonumber\\
&\qquad\qquad\times\int_0^1 \dx u\,(1-u^2)^{-\eps}\left[\frac{1}{u-\beta(s)}-\frac{1}{u+\beta(s)}\right]
\end{align}
with abbreviations
\begin{align}
s_\text{max}&=1-\frac{1-\sqrt{(1-\alpha_3)(1-v_{11})}}{\sqrt{1-(1-\alpha_3)(1-v_{11})}}\,,\\
\alpha_2&=\frac{v_{11} y_2}{1-v_{11}}\,,\\
\alpha_3&=\frac{v_{11} y_3}{1-v_{11}}\,,\\
\beta_3(s)&=\frac{2(1-s)}{s(2-s)}\frac{\sqrt{1-\alpha_3}}{\sqrt{\alpha_3}}\,,\\
s_{\pm_1,\pm_2}&=1\pm_1\frac{1\pm_2 \sqrt{1-\alpha_3}}{\sqrt{\alpha_3}}\,,\\
r_{\pm}&=\frac{\sqrt{1-\alpha_3}\pm\sqrt{1-\alpha_2}}{\sqrt{\alpha_2-\alpha_3}}\,.
\end{align}
At this point it is worth noting that the identity
\begin{align}
\frac{(\alpha_2-\alpha_3)(1-\alpha_2)}{\alpha_3}=\frac{X_{3,3}^2 X_{2,2}^2}{(1-v_{11})Y_3 X_2^2}
\end{align}
holds,
which allows for cancellations between the prefactor that multiplies the branch integral in the full angular integral.
Remaining sign functions that arise in the form $\sqrt{\xi^2}/\xi$ can be dropped by simultaneously dropping the very same $|\dots |$ in the weights $r_\pm$, effectively replacing
\begin{align}
r_\pm\rightarrow\tilde{r}_{\pm}&=\frac{\sqrt{X_2}\sqrt{(1-v_{11})X_3-v_{11}Y_3}\pm X_{2,1}\sqrt{X_3}}{\sqrt{v_{11}}X_{3,1}}\,.
\end{align}
The $u$ integral can be evaluated as a series in $\eps$ in the form
\begin{align}
&\int_0^1 \dx u\,(1-u^2)^{-\eps}\left[\frac{1}{u-\beta(s)}-\frac{1}{u+\beta(s)}\right]
\nonumber\\
&=
\sum_{n=0}^\infty\,(-\eps)^n\,\sum_{\lbrace\pm\rbrace_n}
\left[
\goncharov(\beta(s),\pm 1,\dots,\pm 1;1)-\goncharov(-\beta(s),\pm 1,\dots,\pm 1;1)
\right].
\end{align}
To perform the subsequent $s$ integration, we need to write down the $\goncharov$ functions in the form $G(\dots;s)$ with weights independent of $s$.
A general algorithm for this task has been developed by Panzer \cite{Panzer:2014caa}.
The main two identities are for one \cite{Brown:2009,Brown:2010}
\begin{align}
&\frac{\partial}{\partial x}\,\goncharov(a_1(x),\dots,a_n(x);z)=-\frac{a_1^\prime}{z-a_1}\goncharov(\check{a}_1,\dots;z)\nonumber\\
&
\qquad+\sum_{i=1}^{n-1}\frac{(a_i-a_{i+1})^\prime}{a_i-a_{i+1}}\left[\goncharov(\dots,\check{a}_{i+1},\dots;z)-\goncharov(\dots,\check{a}_{i},\dots;z)\right]-\frac{a_{n}^\prime}{a_n}\,\goncharov(\dots,\check{a}_{n};z)\,,
\label{eq: weight depends on x}
\end{align}
where $z$ does not depend on $x$, the dash means derivative w.r.t. $x$ and $\check{a}_j$ denotes the omission of $a_j$ from the weight vector such that on the right side of eq.\,\eqref{eq: weight depends on x} each $\goncharov$ is of lower weight.
Iterating this differentiation and subsequently `integrating back' removes the $x$ dependence of the weights.
For another, for $x$-independent weights the identity
\begin{align}
\goncharov(a_1,\dots,a_n;z(x))=\int_0^x\frac{\dx x_1}{z(x_1)-a_1}\frac{\partial z(x_1)}{\partial x_1}\,\goncharov(a_2,\dots,a_n;z(x_1))
\end{align}
allows for iterative weight-reduction until $G(z(x^\prime))=1$ and upon re-integration yields a Goncharov polylogarithm with argument $x$ and weights independent of $x$ allowing for algorithmic integration over $x$ according to the defining relation \cite{Goncharov:2001}
\begin{align}
\int_0^z\frac{\dx x_1}{x_1-a_1}\goncharov(a_2,\dots,a_n;x_1)=\goncharov(a_1,a_2,\dots,a_n;z)\,.
\end{align}

In the particular case at hand it is most transparent to proceed in two steps.
First, the integration variable is moved from the weights to the argument in the form
\begin{align}
\goncharov(\beta(s),\pm 1,\dots,\pm 1;1)\longrightarrow\sum_i c_i\,\goncharov\left(\vec{a}_i;\frac{1}{\beta(s)}\right),
\end{align}
with weights $\vec{a}_i$ independent of $s$.
Explicitly the first orders are
\begin{align}
&\int_0^1 \dx u\,(1-u^2)^{-\eps}\left[\frac{1}{u-\beta(s)}-\frac{1}{u+\beta(s)}\right]=
-\goncharov(-1;\beta^{-1})+\goncharov(1;\beta^{-1})\nonumber\\
&\quad+
\eps \left[2 \goncharov(-1,1)
   \goncharov(-1;\beta^{-1})-2 \goncharov(-1,1)
   \goncharov(1;\beta^{-1})+\goncharov(-1,-1;\beta^{-1})   
   -\goncharov(-1,1;\beta^{-1})
\right.\nonumber\\
&\left.   
   \qquad-2
   \goncharov(0,-1;\beta^{-1})+2
   \goncharov(0,1;\beta^{-1})+\goncharov(1,-1;\beta^{-1})-\goncharov(1,1;\beta^{-1})\right]+\mathcal{O}(\eps^2)
\end{align}
Secondly, we perform a change of variables in the last argument, schematically
\begin{align}
\goncharov\left(\vec{a}_i;\frac{1}{\beta(s)}\right)\longrightarrow
\sum_i d_i \goncharov\left(\vec{b}_i;s\right).
\end{align}
Explicitly the first orders are
\begin{align}
&\int_0^1 \dx u\,(1-u^2)^{-\eps}\left[\frac{1}{u-\beta(s)}-\frac{1}{u+\beta(s)}\right]=
\goncharov(s_{--};s)-\goncharov(s_{-+};s)-\goncharov
   (s_{+-};s)+\goncharov(s_{++};s)   
   \nonumber\\
   &+
\eps
   \left[\goncharov(s_{--},s_{-+};s)-\goncharov(s_{-+},s_{--};s)+\goncharov(s_{--},s_{+-};s)-\goncharov(s_{+-},s_{--};s)-\goncharov(s_{--},s_{++};s)
\right.  
    \nonumber\\
   &
   \left.\quad
   -\goncharov(s_{++},s_{--};s)-2 \goncharov(-1,1)
   \goncharov(s_{--};s)+2 \goncharov(0,s_{--};s)+2
   \goncharov(2,s_{--};s)-\goncharov(s_{--},s_{--};s)
   \right.  
    \nonumber\\
   &
   \left.\quad   
   +\goncharov(s_{-+},s_{+-};s)
   +\goncharov(s_{+-},s_{-+};s)-\goncharov(s_{-+},s_{++};s)+\goncharov(s_{++},s_{-+};s)
\right.  
    \nonumber\\
   &
   \left.\quad      
   +2 \goncharov(-1,1)
   \goncharov(s_{-+};s)-2 \goncharov(0,s_{-+};s)-2
   \goncharov(2,s_{-+};s)+\goncharov(s_{-+},s_{-+};s)
   -\goncharov(s_{+-},s_{++};s)
\right.  
    \nonumber\\
   &
   \left.\quad      
   +\goncharov(s_{++},s_{+-};s)+2 \goncharov(-1,1) \goncharov(s_{+-};s)-2
   \goncharov(0,s_{+-};s)-2
   \goncharov(2,s_{+-};s)+\goncharov(s_{+-},s_{+-};s)
  \right.  
    \nonumber\\
   &
   \left.\quad   
   -2 \goncharov(-1,1) \goncharov(s_{++};s)+2
   \goncharov(0,s_{++};s)+2
   \goncharov(2,s_{++};s)-\goncharov(s_{++},s_{++};s)
   \right]+\mathcal{O}(\eps^2)
   \label{eq: Beta integral s in argument}
\end{align}
with 
\begin{align}
s_{\pm_1,\pm_2}=1\pm_1\frac{1\pm_2 \sqrt{1-\alpha_3}}{\sqrt{\alpha_3}}\,.
\end{align}
These weights $s_{\pm\pm}$ also appear in the $(\dots)^\eps$ factor of the $s$ integral.
Having expressed the $u$ integral up to a certain order in $\eps$ in the form of eq.\,\eqref{eq: Beta integral s in argument} where the $s$ dependence of the Goncharov polylogarithms is in their argument and no longer in the weights, performing the remaining $s$-integral is straightforward.
The $(\dots)^\eps$ is expanded in $\eps$ and converted to Goncharov form, then \texttt{PolyLogTools} takes over, applies the shuffle product to bring the integrand into the schematic form
\begin{align}
\sum_i \frac{c_i}{s-(1+\iu r_\pm)}G(\vec{a_i};s)\,.
\end{align}
and finally the $s$-integration is performed.
The first orders read
\begin{align}
&\mathcal{B}_3^{(1)}(\eps)=\mathcal{B}_2^{(1)}(\eps)+2\pi\eps\,\,\frac{\sqrt{\alpha_3}(1-\alpha_3)^\eps}{\sqrt{1-v_{11}}\sqrt{1-\alpha_2}\sqrt{\alpha_2-\alpha_3}}\left(\frac{v_{11}}{1-v_{11}}\right)^{-\eps}
\nonumber\\
&\times
\mathrm{Im}\Big\lbrace
   -\goncharov(1+\iu
   r_-,s_{--};s_\text{max})+\goncharov(1+\iu
   r_-,s_{-+};s_\text{max})+\goncharov(1+\iu
   r_-,s_{+-};s_\text{max})
\nonumber\\
&
   -\goncharov(1+\iu
   r_-,s_{++};s_\text{max})+\goncharov(1+\iu
   r_+,s_{--};s_\text{max})-\goncharov(1+\iu
   r_+,s_{-+};s_\text{max})
   \nonumber\\
&  -\goncharov(1+\iu
   r_+,s_{+-};s_\text{max})+\goncharov(1+\iu
   r_+,s_{++};s_\text{max})
   \nonumber\\
&  +
\eps \Big[-\goncharov(1+\iu
   r_-,s_{--},s_{-+};s_\text{max})+\goncharov(1+\iu
   r_-,s_{-+},s_{--};s_\text{max})
 -\goncharov(1+i
   r_-,s_{--},s_{+-};s_\text{max})
      \nonumber\\
& 
+\goncharov(1+\iu
   r_-,s_{+-},s_{--};s_\text{max})-\goncharov(1+\iu
   r_-,s_{--},s_{++};s_\text{max})-\goncharov(1+\iu
   r_-,s_{++},s_{--};s_\text{max})
   \nonumber\\
&    
   +2 \goncharov(1+\iu
   r_-,s_{--},0;s_\text{max})+2 \goncharov(1+\iu
   r_-,s_{--},2;s_\text{max})-\goncharov(1+\iu
   r_-,s_{--},s_{--};s_\text{max})
   \nonumber\\
&    
   +\goncharov(1+\iu
   r_-,s_{-+},s_{+-};s_\text{max})
   +\goncharov(1+\iu
   r_-,s_{+-},s_{-+};s_\text{max})
   +\goncharov(1+\iu
   r_-,s_{-+},s_{++};s_\text{max})
   \nonumber\\
&    
   -\goncharov(1+\iu
   r_-,s_{++},s_{-+};s_\text{max})-2 \goncharov(1+\iu
   r_-,s_{-+},0;s_\text{max})-2 \goncharov(1+\iu
   r_-,s_{-+},2;s_\text{max})
      \nonumber\\
& +\goncharov(1+\iu
   r_-,s_{-+},s_{-+};s_\text{max})+\goncharov(1+\iu
   r_-,s_{+-},s_{++};s_\text{max})-\goncharov(1+\iu
   r_-,s_{++},s_{+-};s_\text{max})
   \nonumber\\
&    
   -2 \goncharov(1+\iu
   r_-,s_{+-},0;s_\text{max})-2 \goncharov(1+\iu
   r_-,s_{+-},2;s_\text{max})+\goncharov(1+\iu
   r_-,s_{+-},s_{+-};s_\text{max})
   \nonumber\\
&      
   +2 \goncharov(1+\iu
   r_-,s_{++},0;s_\text{max}) 
   +2 \goncharov(1+\iu
   r_-,s_{++},2;s_\text{max})
   -\goncharov(1+\iu
   r_-,s_{++},s_{++};s_\text{max})
   \nonumber\\
&      
   +\goncharov(1+\iu
   r_+,s_{--},s_{-+};s_\text{max})-\goncharov(1+\iu
   r_+,s_{-+},s_{--};s_\text{max})+\goncharov(1+\iu
   r_+,s_{--},s_{+-};s_\text{max})
   \nonumber\\
&      
   -\goncharov(1+\iu
   r_+,s_{+-},s_{--};s_\text{max})+\goncharov(1+\iu
   r_+,s_{--},s_{++};s_\text{max})+\goncharov(1+\iu
   r_+,s_{++},s_{--};s_\text{max})
   \nonumber\\
&      
   -2 \goncharov(1+\iu
   r_+,s_{--},0;s_\text{max})-2 \goncharov(1+\iu
   r_+,s_{--},2;s_\text{max})+\goncharov(1+\iu
   r_+,s_{--},s_{--};s_\text{max})
   \nonumber\\
&      
   -\goncharov(1+\iu
   r_+,s_{-+},s_{+-};s_\text{max})-\goncharov(1+\iu
   r_+,s_{+-},s_{-+};s_\text{max})-\goncharov(1+\iu
   r_+,s_{-+},s_{++};s_\text{max})
   \nonumber\\
&      
   +\goncharov(1+\iu
   r_+,s_{++},s_{-+};s_\text{max})+2 \goncharov(1+\iu
   r_+,s_{-+},0;s_\text{max})+2 \goncharov(1+\iu
   r_+,s_{-+},2;s_\text{max})
   \nonumber\\
&      
   -\goncharov(1+\iu
   r_+,s_{-+},s_{-+};s_\text{max})-\goncharov(1+\iu
   r_+,s_{+-},s_{++};s_\text{max})+\goncharov(1+\iu
   r_+,s_{++},s_{+-};s_\text{max})
   \nonumber\\
&      
   +2 \goncharov(1+\iu
   r_+,s_{+-},0;s_\text{max})+2 \goncharov(1+\iu
   r_+,s_{+-},2;s_\text{max})-\goncharov(1+\iu
   r_+,s_{+-},s_{+-};s_\text{max})
   \nonumber\\
&      
   -2 \goncharov(1+\iu
   r_+,s_{++},0;s_\text{max})-2 \goncharov(1+\iu
   r_+,s_{++},2;s_\text{max})+\goncharov(1+\iu
   r_+,s_{++},s_{++};s_\text{max})\Big]
\nonumber\\
&   
   +\mathcal{O}(\eps^2)
\Big\rbrace\,.
\end{align}
To any order of the expansion, the alphabet is given by the ten letters
\begin{align}
\left\lbrace 0\,, 1\pm_1 \iu r_{\pm_2}\,, s_{\pm_1\pm_2}\,, 2
\right\rbrace.
\end{align}
This branch integral can be used to build up $I_{111}^{(m)}$ to any desired order in $\eps$.
To go to $n=4$ and beyond, one would need to start over before the expansion in $\eps$ and bring $\mathcal{B}_3^{(0/1)}$ into a form that has a simple $\eps$-dependence.
\bibliography{AiMd}

\providecommand{\href}[2]{#2}\begingroup\raggedright\begin{thebibliography}{10}

\bibitem{Ellis:1980}
R.K.~Ellis, M.~Furman, H.~Haber and I.~Hinchliffe, \emph{{Large corrections to
  high-pT hadron-hadron scattering in QCD}},
  \href{https://doi.org/https://doi.org/10.1016/0550-3213(80)90010-3}{\emph{Nucl.
  Phys. B} {\bfseries 173} (1980) }.

\bibitem{Schellekens:1981}
A.~Schellekens, \emph{{Perturbative QCD and lepton pair production}}, Ph.D.
  thesis, Nijmegen University, 6, 1981.

\bibitem{vanNeerven:1985}
W.~van Neerven, \emph{{Dimensional Regularization of Mass and Infrared
  Singularities in Two Loop On-shell Vertex Functions}},
  \href{https://doi.org/10.1016/0550-3213(86)90165-3}{\emph{Nucl. Phys. B}
  {\bfseries 268} (1986) 453}.

\bibitem{Beenakker:1988}
W.~Beenakker, H.~Kuijf, W.~van Neerven and J.~Smith, \emph{{QCD Corrections to
  Heavy Quark Production in p anti-p Collisions}},
  \href{https://doi.org/10.1103/PhysRevD.40.54}{\emph{Phys. Rev. D} {\bfseries
  40} (1989) 54}.

\bibitem{Somogyi:2011}
G.~Somogyi, \emph{{Angular integrals in d dimensions}},
  \href{https://doi.org/10.1063/1.3615515}{\emph{J. Math. Phys.} {\bfseries 52}
  (2011) 083501} [\href{https://arxiv.org/abs/1101.3557}{{\ttfamily
  1101.3557}}].

\bibitem{Lyubovitskij:2021}
{V.\,E. Lyubovitskij, F. Wunder and A.\,S. Zhevlakov}, \emph{{New ideas for
  handling of loop and angular integrals in D-dimensions in QCD}},
  \href{https://doi.org/10.1007/JHEP06(2021)066}{\emph{JHEP} {\bfseries 06}
  (2021) 066} [\href{https://arxiv.org/abs/2102.08943}{{\ttfamily
  2102.08943}}].

\bibitem{Wunder:2024}
F.~Wunder, \emph{Asymptotic behavior of angular integrals in the massless
  limit}, \href{https://doi.org/10.1103/PhysRevD.109.076022}{\emph{Phys. Rev.
  D} {\bfseries 109} (2024) 076022}.

\bibitem{Smirnov:2024pbj}
V.A.~Smirnov and F.~Wunder, \emph{{Expansion by regions meets angular
  integrals}}, \href{https://doi.org/10.1007/JHEP08(2024)138}{\emph{JHEP}
  {\bfseries 08} (2024) 138}
  [\href{https://arxiv.org/abs/2405.13120}{{\ttfamily 2405.13120}}].

\bibitem{Laporta:1994mb}
S.~Laporta, \emph{{Hyperspherical integration and the triple cross vertex
  graphs}}, \href{https://doi.org/10.1007/BF02780705}{\emph{Nuovo Cim. A}
  {\bfseries 107} (1994) 1729}
  [\href{https://arxiv.org/abs/hep-ph/9404203}{{\ttfamily hep-ph/9404203}}].

\bibitem{Haug:2024yfi}
J.~Haug and F.~Wunder, \emph{{Angular integrals with three denominators via
  IBP, mass reduction, dimensional shift, and differential equations}},
  \href{https://doi.org/10.1007/JHEP03(2025)141}{\emph{JHEP} {\bfseries 03}
  (2025) 141} [\href{https://arxiv.org/abs/2410.18177}{{\ttfamily
  2410.18177}}].

\bibitem{Duke:1982}
D.W.~Duke and J.F.~Owens, \emph{Quantum-chromodynamic corrections to
  deep-inelastic compton scattering},
  \href{https://doi.org/10.1103/PhysRevD.26.1600}{\emph{Phys. Rev. D}
  {\bfseries 26} (1982) 1600}.

\bibitem{Hekhorn:2019}
F.~Hekhorn, \emph{{Next-to-Leading Order QCD Corrections to Heavy-Flavour
  Production in Neutral Current DIS}}, Ph.D. thesis, {University of
  T\"ubingen}, 2019.
\newblock \href{https://arxiv.org/abs/1910.01536}{{\ttfamily 1910.01536}}.
\newblock 10.15496/publikation-34811.

\bibitem{Anderle:2016}
D.~Anderle, D.~de~Florian and Y.~Rotstein~Habarnau, \emph{{Towards
  semi-inclusive deep inelastic scattering at next-to-next-to-leading order}},
  \href{https://doi.org/10.1103/PhysRevD.95.034027}{\emph{Phys. Rev. D}
  {\bfseries 95} (2017) 034027}
  [\href{https://arxiv.org/abs/1612.01293}{{\ttfamily 1612.01293}}].

\bibitem{Wang:2019}
B.~Wang, J.~Gonzalez-Hernandez, T.~Rogers and N.~Sato, \emph{{Large Transverse
  Momentum in Semi-Inclusive Deeply Inelastic Scattering Beyond Lowest Order}},
  \href{https://doi.org/10.1103/PhysRevD.99.094029}{\emph{Phys. Rev. D}
  {\bfseries 99} (2019) 094029}
  [\href{https://arxiv.org/abs/1903.01529}{{\ttfamily 1903.01529}}].

\bibitem{Ahmed:2024owh}
T.~Ahmed, S.~Goyal, S.M.~Hasan, R.N.~Lee, S.-O.~Moch, V.~Pathak et~al.,
  \emph{{NNLO phase-space integrals for semi-inclusive deep-inelastic
  scattering}},  \href{https://arxiv.org/abs/2412.16509}{{\ttfamily
  2412.16509}}.

\bibitem{Matsuura:1989}
T.~Matsuura, S.~Van Der~Marck and W.~Van~Neerven, \emph{{The calculation of the
  second order soft and virtual contributions to the Drell-Yan cross section}},
  \href{https://doi.org/https://doi.org/10.1016/0550-3213(89)90620-2}{\emph{Nucl.
  Phys. B} {\bfseries 319} (1989) 570}.

\bibitem{Matsuura:1990}
T.~Matsuura, R.~Hamberg and W.L.~van Neerven, \emph{{The contribution of the
  gluon-gluon subprocess to the Drell-Yan K-factor}},
  \href{https://doi.org/https://doi.org/10.1016/0550-3213(90)90391-P}{\emph{Nucl.
  Phys. B} {\bfseries 345} (1990) 331}.

\bibitem{Hamberg:1991}
R.~Hamberg, W.~Van~Neerven and T.~Matsuura, \emph{{A complete calculation of
  the order $\alpha_s^2$ correction to the Drell-Yan K-factor}},
  \href{https://doi.org/https://doi.org/10.1016/0550-3213(91)90064-5}{\emph{Nucl.
  Phys. B} {\bfseries 359} (1991) 343}.

\bibitem{Mirkes:1992}
E.~Mirkes, \emph{{Angular decay distribution of leptons from W bosons at NLO in
  hadronic collisions}},
  \href{https://doi.org/10.1016/0550-3213(92)90046-E}{\emph{Nucl. Phys. B}
  {\bfseries 387} (1992) 3}.

\bibitem{Bahjat-Abbas:2018}
N.~Bahjat-Abbas, J.~Sinninghe~Damst\'e, L.~Vernazza and C.~White, \emph{{On
  next-to-leading power threshold corrections in Drell-Yan production at
  N$^3$LO}}, \href{https://doi.org/10.1007/JHEP10(2018)144}{\emph{JHEP}
  {\bfseries 10} (2018) 144}
  [\href{https://arxiv.org/abs/1807.09246}{{\ttfamily 1807.09246}}].

\bibitem{Gordon:1993}
L.~Gordon and W.~Vogelsang, \emph{{Polarized and unpolarized prompt photon
  production beyond the leading order}},
  \href{https://doi.org/10.1103/PhysRevD.48.3136}{\emph{Phys. Rev. D}
  {\bfseries 48} (1993) 3136}.

\bibitem{Rein:2024}
D.~Rein, M.~Schlegel and W.~Vogelsang, \emph{Probing the polarized photon
  content of the proton in $ep$ collisions at the eic},
  \href{https://doi.org/10.1103/PhysRevD.110.014041}{\emph{Phys. Rev. D}
  {\bfseries 110} (2024) 014041}
  [\href{https://arxiv.org/abs/2405.04232}{{\ttfamily 2405.04232}}].

\bibitem{Schlegel:2012}
M.~Schlegel, \emph{{Partonic description of the transverse target single-spin
  asymmetry in inclusive deep-inelastic scattering}},
  \href{https://doi.org/10.1103/PhysRevD.87.034006}{\emph{Phys. Rev. D}
  {\bfseries 87} (2013) 034006}
  [\href{https://arxiv.org/abs/1211.3579}{{\ttfamily 1211.3579}}].

\bibitem{Ringer:2015}
F.~Ringer and W.~Vogelsang, \emph{{Single-Spin Asymmetries in W Boson
  Production at Next-to-Leading Order}},
  \href{https://doi.org/10.1103/PhysRevD.91.094033}{\emph{Phys. Rev. D}
  {\bfseries 91} (2015) 094033}
  [\href{https://arxiv.org/abs/1503.07052}{{\ttfamily 1503.07052}}].

\bibitem{Rein:2025qhe}
{Rein, Daniel and Schlegel, Marc and Tollk{\"u}hn, Patrick and Vogelsang,
  Werner}, \emph{{Transverse Nucleon Single-Spin Asymmetry for Single-Inclusive
  Hadron and Jet Production at NLO Accuracy}},
  \href{https://arxiv.org/abs/2503.16119}{{\ttfamily 2503.16119}}.

\bibitem{Rein:2025pwu}
D.~Rein, M.~Schlegel, P.~Tollk{\"u}hn and W.~Vogelsang, \emph{{NLO corrections
  and factorization for transverse single-spin asymmetries}},
  \href{https://arxiv.org/abs/2503.16097}{{\ttfamily 2503.16097}}.

\bibitem{Devoto:2024}
F.~Devoto, K.~Melnikov, R.~R{\"o}ntsch, C.~Signorile-Signorile and
  D.M.~Tagliabue, \emph{{{A fresh look at the nested soft-collinear subtraction
  scheme: NNLO QCD corrections to $N$-gluon final states in $q\bar{q}$
  annihilation}}}, \href{https://doi.org/10.1007/JHEP02(2024)016}{\emph{JHEP}
  {\bfseries 2024} (2024) 1}.

\bibitem{Rowe:2024jml}
M.~Rowe and R.~Zwicky, \emph{{Structure-dependent QED in $ {B}^{-}\to
  {\ell}^{-}\overline{\nu}\left(\gamma \right) $}},
  \href{https://doi.org/10.1007/JHEP07(2024)249}{\emph{JHEP} {\bfseries 07}
  (2024) 249} [\href{https://arxiv.org/abs/2404.07648}{{\ttfamily
  2404.07648}}].

\bibitem{Agarwal:2024gws}
P.~Agarwal, K.~Melnikov and I.~Pedron, \emph{{N-jettiness soft function at
  next-to-next-to-leading order in perturbative QCD}},
  \href{https://doi.org/10.1007/JHEP05(2024)005}{\emph{JHEP} {\bfseries 05}
  (2024) 005} [\href{https://arxiv.org/abs/2403.03078}{{\ttfamily
  2403.03078}}].

\bibitem{Baranowski:2024ysi}
D.~Baranowski, M.~Delto, K.~Melnikov, A.~Pikelner and C.-Y.~Wang, \emph{{Triple
  real-emission contribution to the zero-jettiness soft function at N3LO in
  QCD}},  \href{https://arxiv.org/abs/2412.14001}{{\ttfamily 2412.14001}}.

\bibitem{Baranowski:2025}
D.~Baranowski, M.~Delto, K.~Melnikov, A.~Pikelner and C.-Y.~Wang,
  \emph{Zero-jettiness soft function to third order in perturbative qcd},
  \href{https://doi.org/10.1103/PhysRevLett.134.191902}{\emph{Phys. Rev. Lett.}
  {\bfseries 134} (2025) 191902}.

\bibitem{Lee:2024nqu}
H.~Lee, \emph{{Tensor Integrals in the Large-Scale Structure}},
  \href{https://arxiv.org/abs/2410.13931}{{\ttfamily 2410.13931}}.

\bibitem{tHooft:1972}
G.~{'t Hooft} and M.~Veltman, \emph{Regularization and renormalization of gauge
  fields},
  \href{https://doi.org/https://doi.org/10.1016/0550-3213(72)90279-9}{\emph{Nucl.
  Phys. B} {\bfseries 44} (1972) 189 }.

\bibitem{Bollini:1972}
{C.\,G. Bollini and J.\,J. Giambiagi}, \emph{{Dimensional Renormalization: The
  Number of Dimensions as a Regularizing Parameter}},
  \href{https://doi.org/10.1007/BF02895558}{\emph{{Nuovo Cim. B}} {\bfseries
  12} (1972) 20}.

\bibitem{Salvatori:2024nva}
G.~Salvatori, \emph{{The Tropical Geometry of Subtraction Schemes}},
  \href{https://arxiv.org/abs/2406.14606}{{\ttfamily 2406.14606}}.

\bibitem{Beneke:1997zp}
M.~Beneke and V.A.~Smirnov, \emph{{Asymptotic expansion of Feynman integrals
  near threshold}},
  \href{https://doi.org/10.1016/S0550-3213(98)00138-2}{\emph{Nucl. Phys. B}
  {\bfseries 522} (1998) 321}
  [\href{https://arxiv.org/abs/hep-ph/9711391}{{\ttfamily hep-ph/9711391}}].

\bibitem{Smirnov:2002pj}
V.A.~Smirnov, \emph{{Applied asymptotic expansions in momenta and masses}},
  \href{https://doi.org/10.1007/3-540-44574-9}{\emph{Springer Tracts Mod.
  Phys.} {\bfseries 177} (2002) 1}.

\bibitem{Smirnov:2012gma}
V.A.~Smirnov, \emph{{Analytic tools for Feynman integrals}},
  \href{https://doi.org/10.1007/978-3-642-34886-0}{\emph{Springer Tracts Mod.
  Phys.} {\bfseries 250} (2012) 1}.

\bibitem{Ahmed:2024pxr}
T.~Ahmed, S.M.~Hasan and A.~Rapakoulias, \emph{{Phase-space integrals through
  Mellin-Barnes representation}},
  \href{https://arxiv.org/abs/2410.18886}{{\ttfamily 2410.18886}}.

\bibitem{Tarasov:1996br}
O.V.~Tarasov, \emph{{Connection between Feynman integrals having different
  values of the space-time dimension}},
  \href{https://doi.org/10.1103/PhysRevD.54.6479}{\emph{Phys. Rev. D}
  {\bfseries 54} (1996) 6479}
  [\href{https://arxiv.org/abs/hep-th/9606018}{{\ttfamily hep-th/9606018}}].

\bibitem{Tarasov:2000sf}
O.V.~Tarasov, \emph{{Application and explicit solution of recurrence relations
  with respect to space-time dimension}},
  \href{https://doi.org/10.1016/S0920-5632(00)00849-5}{\emph{Nucl. Phys. B
  Proc. Suppl.} {\bfseries 89} (2000) 237}
  [\href{https://arxiv.org/abs/hep-ph/0102271}{{\ttfamily hep-ph/0102271}}].

\bibitem{Fleischer:2003rm}
J.~Fleischer, F.~Jegerlehner and O.V.~Tarasov, \emph{{A New hypergeometric
  representation of one loop scalar integrals in d dimensions}},
  \href{https://doi.org/10.1016/j.nuclphysb.2003.09.004}{\emph{Nucl. Phys. B}
  {\bfseries 672} (2003) 303}
  [\href{https://arxiv.org/abs/hep-ph/0307113}{{\ttfamily hep-ph/0307113}}].

\bibitem{Davydychev:1990jt}
A.I.~Davydychev, \emph{{Some exact results for N point massive Feynman
  integrals}}, \href{https://doi.org/10.1063/1.529383}{\emph{J. Math. Phys.}
  {\bfseries 32} (1991) 1052}.

\bibitem{Davydychev:1990cq}
A.I.~Davydychev, \emph{{General results for massive N point Feynman diagrams
  with different masses}}, \href{https://doi.org/10.1063/1.529914}{\emph{J.
  Math. Phys.} {\bfseries 33} (1992) 358}.

\bibitem{Anastasiou:1999ui}
C.~Anastasiou, E.W.N.~Glover and C.~Oleari, \emph{{Scalar one loop integrals
  using the negative dimension approach}},
  \href{https://doi.org/10.1016/S0550-3213(99)00637-9}{\emph{Nucl. Phys. B}
  {\bfseries 572} (2000) 307}
  [\href{https://arxiv.org/abs/hep-ph/9907494}{{\ttfamily hep-ph/9907494}}].

\bibitem{Tarasov:2019mqy}
O.V.~Tarasov, \emph{{Functional reduction of Feynman integrals}},
  \href{https://doi.org/10.1007/JHEP02(2019)173}{\emph{JHEP} {\bfseries 02}
  (2019) 173} [\href{https://arxiv.org/abs/1901.09442}{{\ttfamily
  1901.09442}}].

\bibitem{Anastasiou:2002}
C.~Anastasiou and K.~Melnikov, \emph{Higgs boson production at hadron colliders
  in nnlo qcd},
  \href{https://doi.org/10.1016/s0550-3213(02)00837-4}{\emph{Nuclear Physics B}
  {\bfseries 646} (2002) 220–256}.

\bibitem{Anastasiou:2003}
C.~Anastasiou, L.~Dixon, K.~Melnikov and F.~Petriello, \emph{Dilepton rapidity
  distribution in the drell-yan process at next-to-next-to-leading order in
  qcd}, \href{https://doi.org/10.1103/PhysRevLett.91.182002}{\emph{Phys. Rev.
  Lett.} {\bfseries 91} (2003) 182002}.

\bibitem{Anastasiou:2004}
C.~Anastasiou, L.~Dixon, K.~Melnikov and F.~Petriello, \emph{High-precision qcd
  at hadron colliders: Electroweak gauge boson rapidity distributions at
  next-to-next-to leading order},
  \href{https://doi.org/10.1103/PhysRevD.69.094008}{\emph{Phys. Rev. D}
  {\bfseries 69} (2004) 094008}.

\bibitem{Kozlov:2016}
M.G.~Kozlov and R.N.~Lee, \emph{{One-loop pentagon integral in $d$ dimensions
  from differential equations in $\varepsilon$-form}},
  \href{https://doi.org/10.1007/jhep02(2016)021}{\emph{JHEP} {\bfseries 2016}
  (2016) }.

\bibitem{Henn:2022ydo}
J.M.~Henn, A.~Matija{\v{s}}i{\'c} and J.~Miczajka, \emph{{One-loop hexagon
  integral to higher orders in the dimensional regulator}},
  \href{https://doi.org/10.1007/JHEP01(2023)096}{\emph{JHEP} {\bfseries 01}
  (2023) 096} [\href{https://arxiv.org/abs/2210.13505}{{\ttfamily
  2210.13505}}].

\bibitem{Laporta:2000}
S.~Laporta, \emph{{High-precision calculation of multiloop Feynman integrals by
  difference equations}},
  \href{https://doi.org/10.1142/S0217751X00002159}{\emph{Int. J. Mod. Phys. A}
  {\bfseries 15} (2000) 5087}
  [\href{https://arxiv.org/abs/hep-ph/0102033}{{\ttfamily hep-ph/0102033}}].

\bibitem{Lee:2013hzt}
R.N.~Lee and A.A.~Pomeransky, \emph{{Critical points and number of master
  integrals}}, \href{https://doi.org/10.1007/JHEP11(2013)165}{\emph{JHEP}
  {\bfseries 11} (2013) 165} [\href{https://arxiv.org/abs/1308.6676}{{\ttfamily
  1308.6676}}].

\bibitem{Binosi:2003}
D.~Binosi and L.~Theu\ss{}l, \emph{{JaxoDraw: A Graphical user interface for
  drawing Feynman diagrams}},
  \href{https://doi.org/10.1016/j.cpc.2004.05.001}{\emph{Comput. Phys. Commun.}
  {\bfseries 161} (2004) 76}
  [\href{https://arxiv.org/abs/hep-ph/0309015}{{\ttfamily hep-ph/0309015}}].

\bibitem{FIESTA3:2014}
A.~Smirnov, \emph{{FIESTA 3: Cluster-parallelizable multiloop numerical
  calculations in physical regions}},
  \href{https://doi.org/https://doi.org/10.1016/j.cpc.2014.03.015}{\emph{Computer
  Physics Communications} {\bfseries 185} (2014) 2090}.

\bibitem{FIESTA4:2016}
A.~Smirnov, \emph{Fiesta4: Optimized feynman integral calculations with gpu
  support},
  \href{https://doi.org/https://doi.org/10.1016/j.cpc.2016.03.013}{\emph{Computer
  Physics Communications} {\bfseries 204} (2016) 189}.

\bibitem{Smirnov:2021rhf}
A.V.~Smirnov, N.D.~Shapurov and L.I.~Vysotsky, \emph{{FIESTA5: Numerical
  high-performance Feynman integral evaluation}},
  \href{https://doi.org/10.1016/j.cpc.2022.108386}{\emph{Comput. Phys. Commun.}
  {\bfseries 277} (2022) 108386}
  [\href{https://arxiv.org/abs/2110.11660}{{\ttfamily 2110.11660}}].

\bibitem{tHooft:1978}
G.~'t~Hooft and M.J.G.~Veltman, \emph{{Scalar One Loop Integrals}},
  \href{https://doi.org/10.1016/0550-3213(79)90605-9}{\emph{Nucl. Phys. B}
  {\bfseries 153} (1979) 365}.

\bibitem{Buza:1995ie}
M.~Buza, Y.~Matiounine, J.~Smith, R.~Migneron and W.L.~van Neerven,
  \emph{{Heavy quark coefficient functions at asymptotic values Q**2
  {\ensuremath{>}}{\ensuremath{>}} m**2}},
  \href{https://doi.org/10.1016/0550-3213(96)00228-3}{\emph{Nucl. Phys. B}
  {\bfseries 472} (1996) 611}
  [\href{https://arxiv.org/abs/hep-ph/9601302}{{\ttfamily hep-ph/9601302}}].

\bibitem{Blumlein:2020}
J.~Bl\"umlein, A.~De~Freitas, C.~Raab and K.~Sch\"onwald, \emph{{The
  $O(\alpha^2)$ initial state QED corrections to $e^+e^- \rightarrow
  \gamma^*/Z_0^*$}},
  \href{https://doi.org/10.1016/j.nuclphysb.2020.115055}{\emph{Nucl. Phys. B}
  {\bfseries 956} (2020) 115055}
  [\href{https://arxiv.org/abs/2003.14289}{{\ttfamily 2003.14289}}].

\bibitem{Goncharov:2001}
A.B.~Goncharov, \emph{Multiple polylogarithms and mixed tate motives},
  \href{https://arxiv.org/abs/math/0103059}{{\ttfamily math/0103059}}.

\bibitem{Duhr:2019}
C.~Duhr and F.~Dulat, \emph{{PolyLogTools \textemdash{} polylogs for the
  masses}}, \href{https://doi.org/10.1007/JHEP08(2019)135}{\emph{JHEP}
  {\bfseries 08} (2019) 135}
  [\href{https://arxiv.org/abs/1904.07279}{{\ttfamily 1904.07279}}].

\bibitem{Besier:2018}
M.~Besier, D.~Van~Straten and S.~Weinzierl, \emph{{Rationalizing roots: an
  algorithmic approach}},
  \href{https://doi.org/10.4310/CNTP.2019.v13.n2.a1}{\emph{Commun. Num. Theor.
  Phys.} {\bfseries 13} (2019) 253}
  [\href{https://arxiv.org/abs/1809.10983}{{\ttfamily 1809.10983}}].

\bibitem{Panzer:2014caa}
E.~Panzer, \emph{{Algorithms for the symbolic integration of hyperlogarithms
  with applications to Feynman integrals}},
  \href{https://doi.org/10.1016/j.cpc.2014.10.019}{\emph{Comput. Phys. Commun.}
  {\bfseries 188} (2015) 148}
  [\href{https://arxiv.org/abs/1403.3385}{{\ttfamily 1403.3385}}].

\bibitem{Brown:2009}
F.~Brown, \emph{The massless higher-loop two-point function},
  \href{https://doi.org/10.1007/s00220-009-0740-5}{\emph{Communications in
  Mathematical Physics} {\bfseries 287} (2009) 925–958}.

\bibitem{Brown:2010}
F.C.S.~Brown, \emph{{On the periods of some Feynman integrals}},
  \href{https://arxiv.org/abs/0910.0114}{{\ttfamily 0910.0114}}.

\end{thebibliography}\endgroup
\bibliographystyle{JHEP}
\end{document}